\documentclass[showpacs,preprintnumbers]{revtex4}
\usepackage{amsfonts}
\usepackage{amsmath}
\usepackage{graphicx}
\usepackage{dcolumn}
\usepackage{bm}
\usepackage{hyperref}

\input{tcilatex}
\begin{document}

\preprint{}
\title{Atom interferometry in the presence of an external test mass}
\author{B. Dubetsky,$^{\text{1}}$ S. B. Libby,$^{\text{2}}$ P. R. Berman$^{%
\text{3}}$}
\affiliation{$^{\text{1}}$bdubetsky@gmail.com\\
$^{\text{2}}$Physics Division, Physical and Life Sciences, Lawrence
Livermore National Laboratory, Livermore, California 94550, USA\\
$^{\text{3}}$Physics Department, University of Michigan, Ann Arbor, Michigan
48109-1040}
\date{\today }

\begin{abstract}
The influence of an external test mass on the phase of the signal of an atom
interferometer is studied theoretically. Using traditional techniques in
atom optics based on the density matrix equations in the Wigner
representation, we are able to extract the various contributions to the
phase of the signal associated with the classical motion of the atoms, the
quantum correction to this motion resulting from atomic recoil that is
produced when the atoms interact with Raman field pulses, and quantum
corrections to the atomic motion that occur in the time between the Raman
field pulses. By increasing the effective wave vector associated with the
Raman field pulses using modified field parameters, we can increase the
sensitivity of the signal to the point where such quantum corrections can be
measured. The expressions that are derived can be evaluated numerically to
isolate the contribution to the signal from an external test mass. The
regions of validity of the exact and approximate expressions are determined.
\end{abstract}

\maketitle
\preprint{}

\section{Introduction}

Since its birth about 30 years ago \cite{a1}, the field of atom
interferometry (AI) has matured significantly. Experiments based on AI have
been used to measure fundamental constants \cite{a2,a3,a4,a5}, the
acceleration of gravity near the Earth's surface \cite{a6,a6.1,a7,a7.1}, the
gradient of the Earth's gravitational field \cite{a4,a8,a9}, and the
curvature of the gravitational field produced by source masses \cite{a9.1}.
Atom interferometer gyroscopes allow one to measure rotation rates;
experiments have utilized optical fields \cite{a10}, nanofabricated
structures \cite{a11}, and three or four spatially or temporally separated
sets of fields that drive Raman transitions to split and recombine the
matter waves \cite{a12,a13,a14,a15}. The frequency shift arising from a
quadratic Zeeman effect was also measured precisely \cite{a15.1}. There have
been limits set on a non-Newtonian Yukawa-type fifth force \cite{a16} and on
dark energy \cite{a17} using AI, as well as theoretical proposals for using
AI to measure general relativity effects \cite{a18,a19}, including
gravitational waves \cite{a20}. A detailed theoretical analysis of the
combined effect of rotation and gravity on the AI signal has been given \cite%
{a14}, based on three- and four-pulse Raman schemes.

Atom interferometry has also been used to probe the gravitational field
produced by a heavy test mass \cite{a4,a5,a9.1,a16,a17}. Using a
double-difference technique \cite{a4} one can extract that part of the phase
of the AI signal caused by the gravitational field of the test mass. This
article provides a theoretical calculation of this contribution to the
phase, based on an atom interferometer using three Raman field pulses. The
results can be used to optimize measurements of the Newtonian gravitational
constant $G$ and to provide a complete derivation of results outlined in a
previous paper \cite{a21}. Additionally, recently, an analytic,
semi-classical expression for the phase response of an atom interferometer
to an arbitrarily placed, stationary point mass has been derived in \cite%
{a21.1}.

\subsection{Estimated Phase Corrections resulting from the Test Mass}

The phase in an atom interferometer depends on the interactions of the atoms
with the applied Raman fields as well as the motion of the atoms between and
following the applied Raman pulses. The Raman pulses couple two hyperfine
sublevels, $g$\ and $e$, in the atomic ground state manifold and it is the
phase associated with the Raman coherence $\rho _{eg}$\ that is measured
using the interferometer.The presence of a gravitational potential modifies
the atomic trajectories, leading to a modification of the AI phase. It is
this modification of the phase that serves as a measure of the sensitivity
of AI to gravitational effects. Since the Earth's gravitational potential is
only slightly inhomogeneous over the physical extent of the atom
interferometer, it can be approximated by a Taylor series in which only the
lead and gradient terms are retained \cite{a27.1}. Approximate solutions for
the atomic trajectories were obtained in Refs. \cite{a14,c2}, where effects
related to the Earth's rotation (centripetal and Coriolis forces) were also
included. An exact expression for the atom trajectories with these
combination of forces has also been derived for a non-spherical
gravitational source (i.e. for an arbitrary gravity-gradient tensor),
rotating with constant angular velocity \cite{a36}.

\begin{figure}[!t]
\includegraphics[width=10.9cm]{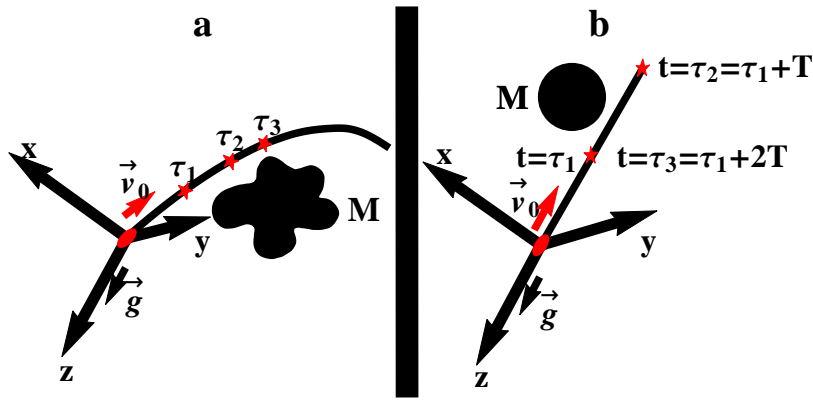}
\caption{Schematic representation of an
atom interferometer in the presence of a test mass $M$. The atom cloud of
the interferometer is launched at $t=0$ with velocity $\mathbf{v}_{0}$ and
interacts with Raman pulses at times $t=\protect\tau _{1}$, $t=\protect\tau %
_{2}\equiv \protect\tau _{1}+T$, and $t=\protect\tau _{3}\equiv \protect\tau %
_{1}+2T$ indicated by the stars in the diagram. (a) A generic
interferometer. (b) The fountain geometry used for the numerical
calculations. In this case the mass is a point mass or spherical mass having
radium $y_{\min }$ that is centered at position $%
(x,y,z)=(x_{m0},y_{m0},z_{m0})$ at time $t=0.$ The case of a stationary test
mass ($x_{m}^{(1)}(t)=0$) and a test mass moving at constant velocity ($%
x_{m}^{(2)}(t)=x_{m0}+v_{m}t$) are considered$.$ The modification of the
signal produced by the test mass would be a maximum if the atom cloud were
to touch the test mass at the top of the cloud's trajectory.}
\label{fig}
\end{figure}

The situation can change dramatically if a massive test object is brought
close to the interferometer (see Fig. \ref{fig}). The accumulated phase
produced by the test mass' gravitational field, $\delta \mathbf{g}\left( 
\mathbf{x,}t\right) ,$ increases with decreasing distance $y_{\min }$
between the test mass and the trajectories of the atoms in the
interferometer and also increases with increasing delay times $T$ between
the Raman pulses. For sufficiently long $T$ and small $y_{\min }$, it is no
longer a good approximation to retain only the lead and gradient terms when
considering the gravitational potential associated with the test mass.

The maximum value of $T$ is limited by experimental considerations; the
largest delay time that has been achieved is $T=1.15$ s \cite{a22}. Even for
smaller delay times, the inhomogeneity of the field can be significant. For
example, with $T=200$ ms, in a symmetric fountain geometry \cite{a23}, the
length of the atomic trajectory is longer than 
\begin{equation}
L=\frac{1}{2}gT^{2}=0.196\text{ m,}  \label{1b}
\end{equation}%
where $\mathbf{g}$ is Earth's gravitational field. With $y_{\min }\lesssim
0.1$ m [see Eqs. (\ref{q}) and (\ref{qq}) in Sec. III], the usually accepted
assumption that the gravitational acceleration is constant or slightly
inhomogeneous along the atom trajectory becomes invalid.

In calculating the atomic trajectories, we can assume \cite{a24} that the
magnitude $\delta g$ of the gravitational field of the test mass at the
position of the atoms in the interferometer is much less than that of the
Earth's field \cite{a24a}, 
\begin{equation}
\delta g\ll g.  \label{2b}
\end{equation}%
Nevertheless, both the average field and field gradient associated with the
test mass can modify the phase of the AI signal. Let us denote the average
field of the test mass over the interferometric path as $\overline{\delta g}$%
. The interferometric phase $\delta \phi $ associated with this average
field strength is of order \cite{a6,a6.1,a7,a7.1} 
\begin{equation}
\delta \phi \sim k\overline{\delta g}T^{2},  \label{1a.1}
\end{equation}%
where $k$ is an effective wave vector of the Raman field and $T$ is the time
delay between Raman pulses. This phase change arises owing to the
acceleration of the atoms produced by the field of the test mass.

In addition to this "classical" contribution to the phase, there are quantum
corrections whose effect we would now like to estimate. Atom interferometers
that make use of copropagating optical fields or copropagating Raman pulses
as their beam splitters and combiners have a signal phase that is
insensitive to quantum corrections if the gravity field is homogeneous.
Quantum corrections arise as a result of rotation \cite{c1} or inhomogeneous
field terms \cite{c2.1,c2}. Quantum corrections $\phi _{q}$ to the phase
from an inhomogeneous gravitational field are of order 
\begin{equation}
\phi _{q}\sim \dfrac{\hbar k^{2}}{M_{a}}\mathbf{\gamma }T^{3},  \label{1a}
\end{equation}%
where $M_{a}$ is an atomic mass and $\mathbf{\gamma }$ is the magnitude of\
the relevant terms in the gravity-gradient tensor. One can understand the
estimate (\ref{1a}) as the quantum part\ of the phase addition $k\mathbf{%
\gamma }vT^{3}$ \cite{c2.1,c2,c2.2} associated with the change of atomic
velocity $v=\hbar k/M_{a}$ owing to recoil \cite{a25} after interaction with
a Raman pulse. When the length of the atomic trajectory $L$ becomes
comparable with the characteristic distance over which the gravitational
potential of the test mass changes, a reasonable estimate for $\mathbf{%
\gamma }$ is $\mathbf{\gamma }\sim \overline{\delta g}/L$. As a consequence,
we find 
\begin{equation}
\dfrac{\phi _{q}}{\delta \phi }\sim \dfrac{\hbar kT}{M_{a}L}.  \label{2a}
\end{equation}%
For Rb$^{\text{87}}$ and $k\approx 1.61\times 10^{7}$ m$^{-1}$ 
\begin{equation}
\dfrac{\phi _{q}}{\delta \phi }\sim 1.2\times 10^{-2}.  \label{2.1a}
\end{equation}%
Calculations \cite{a21} indicate that $\delta \phi $ can be as large as $1$
rad$,$ implying that $\phi _{q}$ can be as large as $10^{-2}$ rad. Since a
lower limit for the phase noise in the interferometer is of order \cite{a22} 
\begin{equation}
\phi _{err}=10^{-3\text{ }}\text{rad,}  \label{2.2a}
\end{equation}%
one sees that quantum corrections $\phi _{q}$ are small but measurable; we
will include them in our considerations.

Another type of quantum correction is produced during the free evolution of
the atomic coherence between the Raman pulses. We formulate the problem in
terms of the Wigner representation \cite{a25.1} for the atomic density
matrix, $\rho \left( \mathbf{x,p,}t\right) $. This is a standard approach
for studying phenomena related to quantization of the atomic center of mass
motion \cite{a25} and laser cooling\ \cite{a25.2}. However, to our
knowledge, it has been used sparingly in the context of AI \cite{a14,a26}.
The convenience of this approach is that, for the time between Raman pulses, 
$\rho \left( \mathbf{x,p,}t\right) $ obeys an equation that is similar to
the classical Liouville equation for the distribution function \cite{a25.1}.

The Wigner distribution function can be written as, 
\begin{subequations}
\label{3}
\begin{eqnarray}
\rho \left( \mathbf{x},\mathbf{p},t\right) &=&\dfrac{1}{\left( 2\pi \hbar
\right) ^{3}}\int d\mathbf{s}\rho _{x}\left( \mathbf{x}+\dfrac{1}{2}\mathbf{s%
},\mathbf{x}-\dfrac{1}{2}\mathbf{s},t\right) \exp \left( -i\mathbf{p}\cdot 
\mathbf{s}/\hbar \right)  \label{3a} \\
&=&\dfrac{1}{\left( 2\pi \hbar \right) ^{3}}\int d\mathbf{u}\rho _{p}\left( 
\mathbf{p}+\dfrac{1}{2}\mathbf{u},\mathbf{p}-\dfrac{1}{2}\mathbf{u},t\right)
\exp \left( i\mathbf{x}\cdot \mathbf{u}/\hbar \right)  \label{3b}
\end{eqnarray}%
where $\rho _{x}\left( \mathbf{x},\mathbf{x}^{\prime },t\right) $ is the
density matrix in the coordinate representation and $\rho _{p}\left( \mathbf{%
p},\mathbf{p}^{\prime },t\right) $ is the density matrix in the momentum
representation. To estimate the quantum corrections we start from the time
evolution equation for the Wigner function for times between the application
of the Raman pulses. In the absence of the Earth's rotation, this equation
can be written as \cite{a14} 
\end{subequations}
\begin{subequations}
\label{4a}
\begin{gather}
\left\{ \partial _{t}+\dfrac{\mathbf{p}}{M_{a}}\partial _{\mathbf{x}%
}-\partial _{\mathbf{x}}U\partial _{\mathbf{p}}+Q\right\} \rho \left( 
\mathbf{x},\mathbf{p},t\right) =0,  \label{4aa} \\
Q=-\left( i\hbar \right) ^{-1}\left[ U\left( \mathbf{x}+\dfrac{1}{2}i\hbar
\partial _{\mathbf{p}}\right) -U\left( \mathbf{x}-\dfrac{1}{2}i\hbar
\partial _{\mathbf{p}}\right) \right] +\partial _{\mathbf{x}}U\partial _{%
\mathbf{p}}  \label{4ba}
\end{gather}%
where $U\left( \mathbf{x}\right) $ is the gravitational potential. For
nearly homogeneous fields, such as the Earth's field, the potential
functions in Eq. (\ref{4ba}) can expanded to first order in $\hbar $. In
that limit, one finds that $Q\sim 0$ and that the Wigner function obeys the
same Liouville equation as the classical density matrix in the time between
pulses. In the presence of a test mass, however, the gravitational potential
is strongly inhomogeneous and higher order terms in the expansion are needed.

Let us estimate the correction from these higher order terms. If the term $%
\partial _{\mathbf{x}}U\partial _{\mathbf{p}}$ in the Liouville equation (%
\ref{4aa}) is responsible for the phase $\delta \phi $ in Eq. (\ref{1a.1})$,$
then the $Q-$term results in a quantum correction 
\end{subequations}
\begin{equation}
\phi _{Q}\sim \dfrac{Q}{\partial _{\mathbf{x}}U\partial _{\mathbf{p}}}\delta
\phi .  \label{4a.1}
\end{equation}%
The density matrix depends on atomic momentum $p$ in two characteristic
ways. There is both a thermal momentum 
\begin{equation}
p_{0}=\sqrt{2M_{a}k_{B}T_{C}}  \label{4a.2}
\end{equation}%
($k_{B}$ is Boltzmann constant, $T_{C}$ is atom cloud temperature) and a
momentum associated with the Doppler phase, 
\begin{equation}
p_{D}\sim \dfrac{M_{a}}{kT}.  \label{4.1a}
\end{equation}%
For Rb at temperature $T_{C}\approx 1\mu K,$ $k=1.61\times 10^{7}$ m$^{-1}$,
and $T=200$ ms,%
\begin{equation}
\dfrac{p_{D}}{p_{0}}=\frac{1}{kv_{0}T}\sim 2.2\times 10^{-5},  \label{5a}
\end{equation}%
where $v_{0}=p_{0}/M_{a}=0.014$ m/s is the thermal velocity. In qualitative
terms, we can think of the dependence of $\rho \left( \mathbf{x},\mathbf{p}%
,t\right) $ on momentum to vary as%
\begin{equation*}
e^{-p^{2}/p_{0}^{2}}e^{ip/p_{D}},
\end{equation*}%
where the first factor gives the thermal distribution and the second a phase
factor resulting from the accumulated Doppler phase between Raman pulses.
Explicit forms for the Doppler phase acquired by the Raman coherence $\rho
_{eg}$\ in a time $T$\ are derived in the next section, but they are
typically of order $kvT=p/p_{D}$. \ 

If $\rho \left( \mathbf{x},\mathbf{p},t\right) \sim
e^{-p^{2}/p_{0}^{2}}e^{ip/p_{D}}$, it follows that the Doppler phase factor
makes the dominant contribution to the momentum gradient since%
\begin{equation}
\partial _{\mathbf{p}}\sim p_{D}^{-1}\gg p_{0}^{-1}.  \label{6a}
\end{equation}%
To estimate the quantum corrections, we expand, Eq. (\ref{4ba}) to second
order in $\hbar $ to obtain%
\begin{equation}
Q\sim \frac{2\hbar ^{2}}{3!}\partial _{\mathbf{x}}^{3}U\left( \frac{\partial
_{\mathbf{p}}}{2}\right) ^{3}.
\end{equation}%
Replacing $\partial _{\mathbf{x}}^{n}U$ by $U/L^{n}$ and $\partial _{\mathbf{%
p}}$ by $p_{D}^{-1}$, we find 
\begin{equation}
\frac{\phi _{Q}}{\delta \phi }\sim \dfrac{Q}{\partial _{\mathbf{x}}U\partial
_{\mathbf{p}}}\sim \dfrac{1}{24}\frac{\partial _{\mathbf{x}}^{2}U}{Up_{D}^{2}%
}=\dfrac{1}{24}\left( \dfrac{\hbar kT}{LM_{a}}\right) ^{2}=5.8\times 10^{-6}.
\label{7a}
\end{equation}%
Consistent with the phase noise given in Eq. (\ref{2.2a}), one should ignore
the $Q-$term in Eq. (\ref{4aa}). However if one uses the AI technique to
measure the Newtonian gravitational constant $G$ with an accuracy of several
ppm (the level achieved is already 150ppm \cite{a5}) then the $Q-$term
should be included. Anticipating innovations capable of reducing the phase
noise to $\phi _{err}\sim 3\times 10^{-7}$rad \cite{a27}, one has to include
the $Q-$term. Consequently, we will include the corrections resulting from
this term.

To summarize, there are two types of quantum corrections to the AI phase
that are to be considered. The first, $\phi _{q}$, arises from inhomogeneous
gravitational field modifications of the Doppler phase associated the recoil
the atoms undergo on interacting with the Raman fields. The ratio $\phi
_{q}/\delta \phi $ is of order $\hbar kT/M_{a}L$. The second, $\phi _{Q}$,
arises from quantum corrections to the off-diagonal elements of the Wigner
distribution during \ periods of free evolution. The ratio $\phi _{Q}/\delta
\phi $ is of order $\left( \hbar kT/M_{a}L\right) ^{2}$.

It is possible to increase both $\delta \phi $ and the quantum corrections $%
\phi _{q}$ and $\phi _{Q}$ using larger values of the effective wave vector $%
k.$ \textbf{Moreover, since }$\delta \phi \propto k$\textbf{, }$\phi
_{q}\propto k^{2}$\textbf{\ and }$\phi _{Q}\propto k^{3}$\textbf{\ [see Eqs.
(\ref{1a.1}, \ref{1a}, \ref{7a})], the relative weight of the quantum
corrections also increases with increasing }$k$\textbf{.} There are at least
five ways to increase $k$: production of higher order atomic density
harmonics in a standing wave field in the Raman-Nath regime [see Eq. (4) in 
\cite{a1}], higher order Bragg scattering \cite{a28}, sequential Bragg
scattering technique \cite{a2}, multicolor techniques \cite{a29}, and Raman
standing wave techniques \cite{a30}. For example, standing wave pulses in
the Raman-Nath regime were used to produce the 10th harmonic of the atomic
density without excessive loss of signal magnitude and without sub-recoil
cooling \cite{a31}. A $4\hbar k$ beam splitter was demonstrated using an
extension of the Raman standing wave technique \cite{a32} and a $51\hbar k$
beam splitter has been produced using higher order Bragg scattering \cite%
{a33}. Recently a high order Bragg scattering atom interferometer was used
to determine the fine structure constant with a resolution 0.25ppb \cite{a3}%
. A $45\hbar k$ beam splitter has been utilized for atom interference using
sequential Bragg scattering \cite{a33.1}. On the theoretical side, it was
shown that, with a proper choice of field polarization, Raman standing waves
in the Raman-Nath regime can be used to create a $4\hbar k$ beam splitter
without increasing the number of separated Raman pulses \cite{a30,a32.1}. To
account for such enhancements, our calculations of the AI's phase are
carried out for an effective $k-$vector that is scaled by an integer factor $%
n_{k}.$

This article is arranged as follows. In the next section we derive exact and
approximate expressions for the phases $\delta \phi ,$ $\phi _{q},$ and $%
\phi _{Q}.$ The results of numerical calculations of the phases are given in
Sec. \ref{s3} for a stationary test mass and a test mass moving at constant
velocity. The calculations enable us to establish the regions of validity of
the approximate expressions for the phases.

\section{\label{s2}Basic Formalism}

The working medium of the atom interferometer consists of a cloud of atoms
that are launched with some initial velocity at $t=0$. The cloud interacts
with three Raman pulses that are separated in time; these pulses couple two
hyperfine sublevels in the atomic ground state manifold. In the time
intervals between the pulses, the atoms move under the influence of a
gravitational potential $U\left( \mathbf{x},t\right) $. The cloud is assumed
to be characterized by a Wigner distribution $f\left( \mathbf{x},\mathbf{p}%
\right) $ at time $t=0$ and the cloud is assumed to be sufficiently
localized such that, at any time, the gravitational field is the same for
all atoms in the cloud. In other words, the cloud can be considered as a
point insofar as its interactions with both the Earth's and the test mass'
gravitational fields. The problem can be broken down into periods of "free
evolution" of density matrix elements before the first Raman pulse is
applied and for the time intervals between subsequent Raman pulses and into
time intervals in which the Raman fields are applied. By "free evolution,"
we mean evolution in the absence of applied radiation fields, but including
the effects produced by $U\left( \mathbf{x},t\right) $. We consider each
region separately and then piece together the total response.

We will see that the quantum corrections leading to $\phi _{q}$ originate in
the recoil the atoms undergo as a result of their interaction with the Raman
pulses. Following the interactions this recoil leads to a contribution to
the Doppler phase of the off-diagonal density matrix elements $\rho _{eg}$ ($%
g$ and $e$ are sublevels of the atoms' ground state manifold) in the time
intervals between the pulses. In addition, the momentum derivatives of the
Doppler phase factors give rise to the $Q-$term corrections; as such, the $%
Q- $term corrections depend only on the free evolution of off-diagonal
density matrix elements between the pulses.

\subsection{Density matrix evolution between the Raman pulses}

Between the Raman pulses, the Wigner function evolves according to Eqs. (\ref%
{4a}). When the distance $L$ over which the gravitational potential energy
varies significantly is much larger than $\hbar $ divided by the
characteristic width $\Delta p$ of the momentum distribution, \textit{i.e.} 
\begin{equation}
\QDABOVE{1pt}{\hbar }{\Delta pL}\ll 1,  \label{5}
\end{equation}%
we can expand $Q$ [Eq. (\ref{4ba})] in a power series in $\hbar $ to obtain 
\begin{equation}
Q\approx -\dfrac{\hbar ^{2}}{24}\chi _{ikl}^{\prime }\left( \mathbf{x}%
,t\right) \partial _{\mathbf{p}_{i}}\partial _{\mathbf{p}_{k}}\partial _{%
\mathbf{p}_{l}},  \label{6aa}
\end{equation}%
where%
\begin{equation}
\chi _{ikl}^{\prime }\left( \mathbf{x},t\right) =-\partial _{x_{i}}\partial
_{x_{k}}\partial _{x_{l}}U\left( \mathbf{x},t\right) .  \label{6b}
\end{equation}%
A summation convention implicit in Eq. (\ref{6aa}) will be used in all
subsequent equations. Repeated indices and symbols appearing on the
right-hand-side (rhs) of an equation are to be summed over, unless they also
appear on the left-hand side (lhs) of that equation.

We have already shown in Eq. (\ref{7a}) that the $Q-$term can be considered
as a small perturbation, allowing us to write 
\begin{equation}
\rho \left( \mathbf{x},\mathbf{p},t\right) =\rho _{0}\left( \mathbf{x},%
\mathbf{p},t\right) +\rho _{Q}\left( \mathbf{x},\mathbf{p},t\right) ,
\label{8}
\end{equation}%
where $\rho _{0}\left( \mathbf{x},\mathbf{p},t\right) $ is the unperturbed
density matrix obeying the equation%
\begin{equation}
\left\{ \partial _{t}+\dfrac{\mathbf{p}}{M_{a}}\partial _{\mathbf{x}%
}-\partial _{\mathbf{x}}U\left( \mathbf{x},t\right) \partial _{\mathbf{p}%
}\right\} \rho _{0}\left( \mathbf{x},\mathbf{p},t\right) =0  \label{9}
\end{equation}%
and $\rho _{Q}\left( \mathbf{x},\mathbf{p},t\right) $ is a perturbation
whose evolution is governed by the equation%
\begin{equation}
\left\{ \partial _{t}+\dfrac{\mathbf{p}}{M_{a}}\partial _{\mathbf{x}%
}-\partial _{\mathbf{x}}U\left( \mathbf{x},t\right) \partial _{\mathbf{p}%
}\right\} \rho _{Q}\left( \mathbf{x},\mathbf{p},t\right) =-Q\rho _{0}\left( 
\mathbf{x},\mathbf{p},t\right) .  \label{10}
\end{equation}%
The $\rho _{0}\left( \mathbf{x},\mathbf{p},t\right) $ term contains the $%
\phi _{q}$ corrections, while the $\rho _{Q}\left( \mathbf{x},\mathbf{p}%
,t\right) $ term provides the $\phi _{Q}$ corrections.

Equation (\ref{9}) has been studied in Ref. \cite{a14} for the Earth's
gravitational field. In this article we obtain a solution of Eq. (\ref{9})
in the presence of a test mass and solve Eq. (\ref{10}) to get the
contribution to the AI phase arising from the $Q-$term. We assume that
density matrix is known at some preceding time $t^{\prime }$ and arbitrarily
set $\rho _{Q}\left( \mathbf{x},\mathbf{p},t^{\prime }\right) =0$ at this
time, such that, at $t=t^{\prime },$ $\rho _{0}\left( \mathbf{x},\mathbf{p}%
,t^{\prime }\right) =\rho \left( \mathbf{x},\mathbf{p},t^{\prime }\right) $.
The solution of the homogeneous Eq. (\ref{9}) is then given by \cite{a14} 
\begin{equation}
\rho _{0}\left( \mathbf{x},\mathbf{p},t\right) =\rho \left[ \mathbf{X}\left( 
\mathbf{x},\mathbf{p},t^{\prime },t\right) ,\mathbf{P}\left( \mathbf{x},%
\mathbf{p},t^{\prime },t\right) ,t^{\prime }\right] ,  \label{12}
\end{equation}%
where $\left\{ \mathbf{X}\left( \mathbf{x},\mathbf{p},t_{1},t_{2}\right) ,%
\mathbf{P}\left( \mathbf{x},\mathbf{p},t_{1},t_{2}\right) \right\} $ are
atomic classical position and momentum at time $t_{1}$ subject to the
constraint that the position and momentum are specified by $\left\{ \mathbf{x%
},\mathbf{p}\right\} $ at time $t_{2}.$ In other words, in Eq. (\ref{12}),
we look for the values $\left\{ \mathbf{X}\left( \mathbf{x},\mathbf{p}%
,t^{\prime },t\right) ,\mathbf{P}\left( \mathbf{x},\mathbf{p},t^{\prime
},t\right) \right\} $ for which $\left\{ \mathbf{X}\left( t^{\prime }\right)
,\mathbf{P}\left( t^{\prime }\right) \right\} $ will lead to values $\left\{ 
\mathbf{X}\left( t\right) ,\mathbf{P}\left( t\right) \right\} =\left\{ 
\mathbf{x},\mathbf{p}\right\} $ under the influence of the applied fields.

Turning our attention to Eq. (\ref{10}), we see that the curly brackets in
that equation is a total time derivative, enabling us to write%
\begin{equation}
\frac{d\rho _{Q}\left( \mathbf{x},\mathbf{p},t^{\prime \prime }\right) }{%
dt^{\prime \prime }}=-Q\rho _{0}\left( \mathbf{x},\mathbf{p},t^{\prime
\prime }\right)  \label{12a}
\end{equation}%
Integrating this equation from $t^{\prime \prime }$ equals $t^{\prime }$ to $%
t,$ using the fact that $\rho _{Q}\left( \mathbf{x},\mathbf{p},t^{\prime
}\right) =0$, and making use of Eqs. (\ref{12}), (\ref{6aa}), and (\ref{6b})$%
,$ we find%
\begin{equation}
\rho _{Q}\left( \mathbf{x},\mathbf{p},t\right) =\dfrac{\hbar ^{2}}{24}%
\int_{t^{\prime }}^{t}dt^{\prime \prime }\left[ \chi _{ikl}^{\prime }\left( 
\mathbf{\xi },t^{\prime \prime }\right) \partial _{\mathbf{\pi }%
_{i}}\partial _{\mathbf{\pi }_{k}}\partial _{\mathbf{\pi }_{l}}\rho
_{0}\left( \mathbf{\xi },\mathbf{\pi },t^{\prime \prime }\right) \right] _{%
\mathbf{\xi }=\mathbf{X}\left( \mathbf{x},\mathbf{p},t^{\prime \prime
},t\right) ,\mathbf{\pi }=\mathbf{P}\left( \mathbf{x},\mathbf{p},t^{\prime
\prime },t\right) }.  \label{15.1}
\end{equation}%
Using Eq. (\ref{12}) one more time, we arrive at%
\begin{equation}
\rho _{Q}\left( \mathbf{x},\mathbf{p},t\right) =\dfrac{\hbar ^{2}}{24}%
\int_{t^{\prime }}^{t}dt^{\prime \prime }\left[ \chi _{ikl}^{\prime }\left( 
\mathbf{\xi },t^{\prime \prime }\right) \partial _{\mathbf{\pi }%
_{i}}\partial _{\mathbf{\pi }_{k}}\partial _{\mathbf{\pi }_{l}}\rho
_{0}\left( \mathbf{X}\left( \mathbf{\xi },\mathbf{\pi },t^{\prime
},t^{\prime \prime }\right) ,\mathbf{P}\left( \mathbf{\xi },\mathbf{\pi }%
,t^{\prime },t^{\prime \prime }\right) ,t^{\prime }\right) \right] _{\left\{ 
\begin{array}{c}
_{\mathbf{\xi }} \\ 
_{\mathbf{\pi }}%
\end{array}%
\right\} =\left\{ 
\begin{array}{c}
\mathbf{X}\left( \mathbf{x},\mathbf{p},t^{\prime \prime },t\right) \\ 
\mathbf{P}\left( \mathbf{x},\mathbf{p},t^{\prime \prime },t\right)%
\end{array}%
\right\} }.  \label{15}
\end{equation}

\subsection{Changes in Density Matrix Elements Produced by the Raman Pulses}

Consider now a cloud of atoms having two hyperfine sublevels $g$ and $e$ in
the ground state manifold. The atoms are prepared in level $g$ at time $t=0$
and they proceed to interact with a $\dfrac{\pi }{2}-\pi -\dfrac{\pi }{2}$
sequence of Raman pulses applied at times%
\begin{equation}
\tau =\left\{ \tau _{1},\tau _{2}=\tau _{1}+T,\tau _{3}=\tau _{1}+2T\right\}
,  \label{16}
\end{equation}%
where $\tau _{1}$ is time delay between cloud launch and the first Raman
pulse and $T$ is the time delay between pulses. The initial atomic density
matrix (\ref{3}) is given by 
\begin{subequations}
\label{17}
\begin{gather}
\rho _{gg}\left( \mathbf{x},\mathbf{p},0\right) =f\left( \mathbf{x},\mathbf{p%
}\right) ,  \label{17a} \\
\rho _{eg}\left( \mathbf{x},\mathbf{p},0\right) =\rho _{ee}\left( \mathbf{x},%
\mathbf{p},0\right) =0,  \label{17b}
\end{gather}%
where $f\left( \mathbf{x},\mathbf{p}\right) $ is the Wigner distribution at $%
t=0.$

If a $\pi /2-$pulse applied at time $\tau _{j},$ the density matrix elements
undergo changes given by \cite{a14} 
\end{subequations}
\begin{subequations}
\label{18}
\begin{gather}
\rho _{ee}\left( \mathbf{x},\mathbf{p},\tau _{j+}\right) =\dfrac{1}{2}\left[
\rho _{ee}\left( \mathbf{x},\mathbf{p},\tau _{j-}\right) +\rho _{gg}\left( 
\mathbf{x},\mathbf{p}-\hbar \mathbf{k},\tau _{j-}\right) \right] +\func{Re}%
\left\{ i\exp \left[ -i\left( \mathbf{k}\cdot \mathbf{x}-\delta
_{12}^{(j)}\tau _{j}-\phi _{j}\right) \right] \rho _{eg}\left( \mathbf{x},%
\mathbf{p}-\dfrac{\hbar \mathbf{k}}{2},\tau _{j-}\right) \right\} ,
\label{18a} \\
\rho _{gg}\left( \mathbf{x},\mathbf{p},\tau _{j+}\right) =\dfrac{1}{2}\left[
\rho _{ee}\left( \mathbf{x},\mathbf{p}+\hbar \mathbf{k},\tau _{j-}\right)
+\rho _{gg}\left( \mathbf{x},\mathbf{p},\tau _{j-}\right) \right] -\func{Re}%
\left\{ i\exp \left[ -i\left( \mathbf{k}\cdot \mathbf{x}-\delta
_{12}^{(j)}\tau _{j}-\phi _{j}\right) \right] \rho _{eg}\left( \mathbf{x},%
\mathbf{p}+\dfrac{\hbar \mathbf{k}}{2},\tau _{j-}\right) \right\} ,
\label{18b} \\
\rho _{eg}\left( \mathbf{x},\mathbf{p},\tau _{j+}\right) =\dfrac{i}{2}\exp %
\left[ i\left( \mathbf{k}\cdot \mathbf{x}-\delta _{12}^{(j)}\tau _{j}-\phi
_{j}\right) \right] \left[ \rho _{ee}\left( \mathbf{x},\mathbf{p}+\dfrac{%
\hbar \mathbf{k}}{2},\tau _{j-}\right) -\rho _{gg}\left( \mathbf{x},\mathbf{p%
}-\dfrac{\hbar \mathbf{k}}{2},\tau _{j-}\right) \right]  \notag \\
+\dfrac{1}{2}\left\{ \rho _{eg}\left( \mathbf{x},\mathbf{p},\tau
_{j-}\right) +\exp \left[ 2i\left( \mathbf{k}\cdot \mathbf{x}-\delta
_{12}^{(j)}\tau _{j}-\phi _{j}\right) \right] \rho _{ge}\left( \mathbf{x},%
\mathbf{p},\tau _{j-}\right) \right\} ,.  \label{18c}
\end{gather}%
Similarly, for $\pi -$pulse applied at time $\tau _{j}$, 
\end{subequations}
\begin{subequations}
\label{19}
\begin{gather}
\rho _{ee}\left( \mathbf{x},\mathbf{p},\tau _{j+}\right) =\rho _{gg}\left( 
\mathbf{x},\mathbf{p}-\hbar \mathbf{k},\tau _{j-}\right) ,  \label{19a} \\
\rho _{gg}\left( \mathbf{x},\mathbf{p},\tau _{j+}\right) =\rho _{ee}\left( 
\mathbf{x},\mathbf{p}+\hbar \mathbf{k},\tau _{j-}\right) ,  \label{19b} \\
\rho _{eg}\left( \mathbf{x},\mathbf{p},\tau _{j+}\right) =\exp \left[
2i\left( \mathbf{k\cdot x}-\delta _{12}^{(j)}\tau _{j}-\phi _{j}\right) %
\right] \rho _{ge}\left( \mathbf{x},\mathbf{p},\tau _{j-}\right) ,
\label{19c}
\end{gather}%
In these equations, $\mathbf{k}$ is an effective wave vector (assumed to be
the same for all the pulses), $\delta _{12}^{(j)}$ is the detuning between
the hyperfine transition frequency and the effective frequency of the Raman
fields (that is the frequency difference of the two fields used to create
the Raman pulse), $\phi _{j}$ is the phase difference between traveling
components of the Raman field, and $\tau _{j\pm }$ are times just after and
before the pulse. We allow pulses to have different detunings and phases $%
\delta _{12}^{\left( j\right) },\phi _{j}$ ($j=1,2,3$).

It is assumed that the temporal width of the Raman pulses are sufficiently
short to guarantee that all phases related to the detuning, Doppler shifts,
and the gravitational fields are effectively frozen during the application\
of the pulses. In addition we assume that the Raman field amplitude and
phase are constant over the size of the atomic cloud, allowing us to neglect
corrections arising from the ac-Stark effect and wave front curvature of the
Raman fields. In principle most of these assumptions are not necessary. One
can derive and explore the analogue of Eqs. (\ref{18}, \ref{19})\
considering extended atom clouds at finite temperature, including
corrections arising from Doppler broadening, ac-Stark effects and
gravitational acceleration produced \textit{during} the Raman pulses. In
this case, however, the corrections depend on the initial atomic
distribution $f\left( \mathbf{x},\mathbf{p}\right) .$ Since this
distribution is usually not known accurately, it is preferable for high
precision atomic interferometry to use Raman pulses of sufficiently short
duration, sufficiently large diameter and sufficiently flat wave fronts to
avoid such corrections.

If a $\pi /2$ pulse acts on a ground state atom, it produces a superposition
of ground and excited states. If there was a momentum $\mathbf{p}$
associated with the ground state amplitude $a_{g}(\mathbf{p})$ before the
pulse is applied, the excited state amplitude $a_{e}(\mathbf{p})$ depends on 
$a_{g}(\mathbf{p-}\hbar \mathbf{k})$. As a consequence, the off-diagonal
density matrix element following the pulse involves the product of state
amplitudes evolving with different momenta. It is this difference in
momentum that leads to the $Q-$term correction in periods of free evolution.

\subsection{AI Signal}

Our goal is to calculate $\rho _{ee}\left( \mathbf{x},\mathbf{p},\tau
_{3+}\right) $, the excited state atomic density matrix element following
the 3rd Raman pulse, since $\rho _{ee}\left( \mathbf{x},\mathbf{p},\tau
_{3+}\right) $ can be related to experimentally measurable quantities. To
carry out the calculation, we use Eqs. (\ref{12}, \ref{15}) for the "free
evolution" of density matrix elements before the first Raman pulse is
applied and for the time intervals between subsequent Raman pulses and use
Eqs. (\ref{18}, \ref{19}) for changes in the density matrix elements
resulting from the application of the Raman pulses. In these free evolution
regions, density matrix elements are affected by the presence of a
gravitational potential that ultimately contributes to the phase of the AI
signal.

From the time the cloud is launched at $t=0$ to the time $\tau _{1}$ that
the first Raman pulse is applied, the only non-vanishing density matrix
element is $\rho _{gg}\left( \mathbf{x},\mathbf{p},t\right) $. In the time
interval between $t=0$ and $t=\tau _{1}$, this density matrix element
evolves to 
\end{subequations}
\begin{equation}
\rho _{gg}\left( \mathbf{x},\mathbf{p},\tau _{1-}\right) =f\left( \mathbf{X}%
\left( \mathbf{x},\mathbf{p},0,\tau _{1}\right) ,\mathbf{P}\left( \mathbf{x},%
\mathbf{p},0,\tau _{1}\right) \right) .  \label{20}
\end{equation}%
For reasons to be discussed below, corrections from the $Q$ term can be
neglected in this time interval. After the first $\pi /2-$pulse, the density
matrix elements change to 
\begin{subequations}
\label{21}
\begin{eqnarray}
\rho _{ee}\left( \mathbf{x},\mathbf{p},\tau _{1+}\right) &=&\frac{1}{2}%
f\left( \mathbf{X}\left( \mathbf{x},\mathbf{p}-\hbar \mathbf{k},0,\tau
_{1}\right) ,\mathbf{P}\left( \mathbf{x},\mathbf{p}-\hbar \mathbf{k},0,\tau
_{1}\right) \right) ,  \label{21a} \\
\rho _{gg}\left( \mathbf{x},\mathbf{p},\tau _{1+}\right) &=&\frac{1}{2}%
f\left( \mathbf{X}\left( \mathbf{x},\mathbf{p},0,\tau _{1}\right) ,\mathbf{P}%
\left( \mathbf{x},\mathbf{p},0,\tau _{1}\right) \right) ,  \label{21b} \\
\rho _{eg}\left( \mathbf{x},\mathbf{p},\tau _{1+}\right) &=&-\dfrac{i}{2}%
\exp \left[ i\left( \mathbf{k}\cdot \mathbf{x}-\delta _{12}^{\left( 1\right)
}\tau _{1}-\phi _{1}\right) \right] f\left( \mathbf{X}\left( \mathbf{x},%
\mathbf{p}-\frac{\hbar \mathbf{k}}{2},0,\tau _{1}\right) ,\mathbf{P}\left( 
\mathbf{x},\mathbf{p}-\frac{\hbar \mathbf{k}}{2},0,\tau _{1}\right) \right) .
\label{21c}
\end{eqnarray}%
One uses these density matrix elements as initial values for the free
evolution between the 1st and 2nd pulses of the unperturbed density matrix;
that is, 
\end{subequations}
\begin{equation}
\rho _{0}\left( \mathbf{x},\mathbf{p},\tau _{1+}\right) =\rho \left( \mathbf{%
x},\mathbf{p},\tau _{1+}\right) .  \label{21.1}
\end{equation}

We now consider the modifications produced by the $Q-$term in the time
interval between the 1st and 2nd pulses. The modifications produced by the $%
Q-$term (\ref{15}) in the atomic coherence before the second pulse acts, $%
\rho _{Qeg}$ can be calculated from Eqs. (\ref{15}, \ref{21c}, \ref{21.1})
as 
\begin{gather}
\rho _{Qeg}\left( \mathbf{x},\mathbf{p},\tau _{2-}\right) =-i\dfrac{\hbar
^{2}}{48}\int_{\tau _{1}}^{\tau _{2}}dt  \notag \\
\times \left\{ \chi _{ikl}^{\prime }\left( \mathbf{\xi },t\right) \partial _{%
\mathbf{\pi }_{i}}\partial _{\mathbf{\pi }_{k}}\partial _{\mathbf{\pi }_{l}}%
\left[ 
\begin{array}{c}
\exp \left[ i\left( \mathbf{k}\cdot \mathbf{X}\left( \mathbf{\xi },\mathbf{%
\pi },\tau _{1},t\right) -\delta _{12}^{\left( 1\right) }\tau _{1}-\phi
_{1}\right) \right] \\ 
\times f\left( 
\begin{array}{c}
\mathbf{X}\left( \mathbf{X}\left( \mathbf{\xi },\mathbf{\pi },\tau
_{1},t\right) ,\mathbf{P}\left( \mathbf{\xi },\mathbf{\pi },\tau
_{1},t\right) -\frac{\hbar \mathbf{k}}{2},0,\tau _{1}\right) , \\ 
\mathbf{P}\left( \mathbf{X}\left( \mathbf{\xi },\mathbf{\pi },\tau
_{1},t\right) ,\mathbf{P}\left( \mathbf{\xi },\mathbf{\pi },\tau
_{1},t\right) -\frac{\hbar \mathbf{k}}{2},0,\tau _{1}\right)%
\end{array}%
\right)%
\end{array}%
\right] \right\} _{\left\{ 
\begin{array}{c}
_{\mathbf{\xi }} \\ 
_{\mathbf{\pi }}%
\end{array}%
\right\} =\left\{ 
\begin{array}{c}
_{\mathbf{X}} \\ 
_{\mathbf{P}}%
\end{array}%
\right\} \left( \mathbf{x},\mathbf{p},t,\tau _{2}\right) }.  \label{22}
\end{gather}%
In Eq. (\ref{22}), the $\mathbf{\pi }$ derivatives lead to two types of
terms. The first of these originates from the thermal distribution and is of
order 
\begin{equation}
\partial _{\mathbf{\pi }_{i}\text{Thermal}}\sim p_{0}^{-1},  \label{23}
\end{equation}%
where $p_{0}$ is thermal momentum defined in Eq. (\ref{4a.2}). The second
arises from the phase factor $\exp \left[ i\left( \mathbf{k}\cdot \mathbf{X}%
\left( \mathbf{\xi },\mathbf{\pi },\tau _{1},t\right) -\delta _{12}^{\left(
1\right) }\tau _{1}-\phi _{1}\right) \right] $ in Eq. (\ref{22}), evaluated
at $t-\tau _{1}\sim T.$ To estimate this contribution, we "turn off" the
gravitational field. In this approximation%
\begin{equation}
\mathbf{X}\left( \mathbf{\xi },\mathbf{\pi },\tau _{1},t\right) =\mathbf{\xi 
}-\mathbf{\pi }\left( t-\tau _{1}\right) /M_{a}  \label{25}
\end{equation}%
and the Doppler phase becomes equal to $\mathbf{k}\cdot \mathbf{\pi }\left(
t-\tau _{1}\right) /M_{a}.$ This phase factor is a rapidly oscillating
function of momentum $\mathbf{\pi }$ having period of order $p_{D}$ defined
by Eq. (\ref{4.1a})$,$ from which we find 
\begin{equation}
\partial _{\mathbf{\pi }_{i}\text{Doppler}}\sim p_{D}^{-1}.  \label{27}
\end{equation}%
In the limit that%
\begin{equation}
kv_{0}T\gg 1,  \label{28}
\end{equation}%
the thermal derivative is smaller than Doppler derivative by the ratio given
in Eq. (\ref{5a}) and can be neglected.

When inequality (\ref{28}) holds, the time separation between pulses $T$ is
sufficiently large to insure that the dominant contributions to the $Q-$
term comes from the momentum derivatives of the Doppler phase factor. As we
will show, the atomic levels' populations ($\rho _{ee}$ and $\rho _{gg}$)
have no phase factor for $0<t<\tau _{3-};$ therefore the $Q-$term
corrections arise only from the atomic coherence $\rho _{eg}.~$As a
consequence, we can neglect any contribution to the $Q-$ term corrections
from atomic state populations. It was for this reason we did not include any 
$Q-$term corrections to the Wigner distribution for the time interval $%
0<t<\tau _{1-}$. \textit{In the Doppler limit defined by Eq. (\ref{28}), the
AI phase is pretty much independent of the atomic momentum and spatial
distributions.}

Calculating the derivatives and retaining those contributions to the
derivatives arising from the Doppler phase only, we arrive at%
\begin{gather}
\rho _{Qeg}\left( \mathbf{x},\mathbf{p},\tau _{2-}\right) =-\dfrac{\hbar ^{2}%
}{48}k_{u}k_{v}k_{w}\int_{\tau _{1}}^{\tau _{2}}dt  \notag \\
\times \left[ \chi _{ikl}^{\prime }\left( \mathbf{\xi },t\right) \partial _{%
\mathbf{\pi }_{i}}\mathbf{X}_{u}\left( \mathbf{\xi },\mathbf{\pi },\tau
_{1},t\right) \partial _{\mathbf{\pi }_{k}}\mathbf{X}_{v}\left( \mathbf{\xi }%
,\mathbf{\pi },\tau _{1},t\right) \partial _{\mathbf{\pi }_{l}}\mathbf{X}%
_{w}\left( \mathbf{\xi },\mathbf{\pi },\tau _{1},t\right) \right] _{\left\{ 
\begin{array}{c}
_{\mathbf{\xi }} \\ 
_{\mathbf{\pi }}%
\end{array}%
\right\} =\left\{ 
\begin{array}{c}
_{\mathbf{X}} \\ 
_{\mathbf{P}}%
\end{array}%
\right\} \left( \mathbf{x},\mathbf{p},t,\tau _{2}\right) }  \notag \\
\left\{ 
\begin{array}{c}
\exp \left[ i\left( \mathbf{k}\cdot \mathbf{X}\left( \mathbf{\xi },\mathbf{%
\pi },\tau _{1},t\right) -\delta _{12}^{\left( 1\right) }\tau _{1}-\phi
_{1}\right) \right] \\ 
\times f\left( 
\begin{array}{c}
\mathbf{X}\left( \mathbf{X}\left( \mathbf{\xi },\mathbf{\pi },\tau
_{1},t\right) ,\mathbf{P}\left( \mathbf{\xi },\mathbf{\pi },\tau
_{1},t\right) -\frac{\hbar \mathbf{k}}{2},0,\tau _{1}\right) , \\ 
\mathbf{P}\left( \mathbf{X}\left( \mathbf{\xi },\mathbf{\pi },\tau
_{1},t\right) ,\mathbf{P}\left( \mathbf{\xi },\mathbf{\pi },\tau
_{1},t\right) -\frac{\hbar \mathbf{k}}{2},0,\tau _{1}\right)%
\end{array}%
\right)%
\end{array}%
\right\} _{\left\{ 
\begin{array}{c}
_{\mathbf{\xi }} \\ 
_{\mathbf{\pi }}%
\end{array}%
\right\} =\left\{ 
\begin{array}{c}
_{\mathbf{X}} \\ 
_{\mathbf{P}}%
\end{array}%
\right\} \left( \mathbf{x},\mathbf{p},t,\tau _{2}\right) },  \label{30}
\end{gather}%
where $k_{u}$ is the $u$th component of the effective $k-$vector. In this
approximation, the derivative no longer acts on the term inside the curly
brackets. Therefore we can apply the multiplication law, 
\begin{equation}
\left\{ 
\begin{array}{c}
\mathbf{X} \\ 
\mathbf{P}%
\end{array}%
\right\} \left( \mathbf{X}\left( \mathbf{x},\mathbf{p},t^{\prime },t^{\prime
\prime }\right) ,\mathbf{P}\left( \mathbf{x},\mathbf{p},t^{\prime
},t^{\prime \prime }\right) ,t,t^{\prime }\right) =\left\{ 
\begin{array}{c}
\mathbf{X} \\ 
\mathbf{P}%
\end{array}%
\right\} \left( \mathbf{x},\mathbf{p},t,t^{\prime \prime }\right)  \label{31}
\end{equation}%
to get%
\begin{equation}
\left\{ 
\begin{array}{c}
\mathbf{X} \\ 
\mathbf{P}%
\end{array}%
\right\} \left( \mathbf{\xi },\mathbf{\pi },\tau _{1},t\right) _{\mathbf{\xi 
}=\mathbf{X}\left( \mathbf{x},\mathbf{p},t,\tau _{2}\right) ,\mathbf{\pi }=%
\mathbf{P}\left( \mathbf{x},\mathbf{p},t,\tau _{2}\right) }=\left\{ 
\begin{array}{c}
\mathbf{X} \\ 
\mathbf{P}%
\end{array}%
\right\} \left( \mathbf{x},\mathbf{p},\tau _{1},\tau _{2}\right) .
\label{32}
\end{equation}%
The expression inside the curly brackets of Eq. (\ref{30}) becomes $t-$%
independent and the $Q-$term just before the second pulse is given by%
\begin{gather}
\rho _{Qeg}\left( \mathbf{x},\mathbf{p},\tau _{2-}\right) =-\dfrac{\hbar ^{2}%
}{48}\left\{ \exp \left[ i\left( \mathbf{k}\cdot \mathbf{\xi }-\delta
_{12}^{\left( 1\right) }\tau _{1}-\phi _{1}\right) \right] f\left( \mathbf{%
\xi },\mathbf{\pi }\right) \right\} _{\left\{ 
\begin{array}{c}
_{\mathbf{\xi }} \\ 
_{\mathbf{\pi }}%
\end{array}%
\right\} =\left\{ 
\begin{array}{c}
_{\mathbf{X}\left( \mathbf{x},\mathbf{p},\tau _{1},\tau _{2}\right) } \\ 
_{\mathbf{P}\left( \mathbf{x},\mathbf{p},\tau _{1},\tau _{2}\right) -\hbar 
\mathbf{k}/2}%
\end{array}%
\right\} }  \notag \\
\times k_{u}k_{v}k_{w}\int_{\tau _{1}}^{\tau _{2}}dt\left[ \chi
_{ikl}^{\prime }\left( \mathbf{\xi },t\right) \partial _{\mathbf{\pi }_{i}}%
\mathbf{X}_{u}\left( \mathbf{\xi },\mathbf{\pi },\tau _{1},t\right) \partial
_{\mathbf{\pi }_{k}}\mathbf{X}_{v}\left( \mathbf{\xi },\mathbf{\pi },\tau
_{1},t\right) \partial _{\mathbf{\pi }_{l}}\mathbf{X}_{w}\left( \mathbf{\xi }%
,\mathbf{\pi },\tau _{1},t\right) \right] _{\left\{ 
\begin{array}{c}
_{\mathbf{\xi }} \\ 
_{\mathbf{\pi }}%
\end{array}%
\right\} =\left\{ 
\begin{array}{c}
_{\mathbf{X}} \\ 
_{\mathbf{P}}%
\end{array}%
\right\} \left( \mathbf{x},\mathbf{p},t,\tau _{2}\right) }.  \label{33}
\end{gather}

We still need an expression for the time evolution of $\rho _{0}\left( 
\mathbf{x},\mathbf{p},t\right) $ between the first and second pulses. From
Eqs. (\ref{12}, \ref{21}, \ref{31}), we find 
\begin{subequations}
\label{34}
\begin{gather}
\rho _{ee}\left( \mathbf{x},\mathbf{p},\tau _{2-}\right) =\frac{1}{2}f\left( 
\mathbf{X}\left( \mathbf{\xi },\mathbf{\pi },0,\tau _{1}\right) ,\mathbf{P}%
\left( \mathbf{\xi },\mathbf{\pi },0,\tau _{1}\right) \right) _{\mathbf{\xi }%
=\mathbf{X}\left( \mathbf{x},\mathbf{p},\tau _{1},\tau _{2}\right) ,\mathbf{%
\pi }=\mathbf{P}\left( \mathbf{x},\mathbf{p},\tau _{1},\tau _{2}\right)
-\hbar \mathbf{k}},  \label{34a} \\
\rho _{gg}\left( \mathbf{x},\mathbf{p},\tau _{2-}\right) =\frac{1}{2}f\left( 
\mathbf{X}\left( \mathbf{x},\mathbf{p},0,\tau _{2}\right) ,\mathbf{P}\left( 
\mathbf{x},\mathbf{p},0,\tau _{2}\right) \right) ,  \label{34b} \\
\rho _{0eg}\left( \mathbf{x},\mathbf{p},\tau _{2-}\right) =-\dfrac{i}{2}%
\left\{ \exp \left[ i\left( \mathbf{k}\cdot \mathbf{\xi }-\delta
_{12}^{\left( 1\right) }\tau _{1}-\phi _{1}\right) \right] \right.  \notag \\
\left. \times f\left( \mathbf{X}\left( \mathbf{\xi },\mathbf{\pi },0,\tau
_{1}\right) ,\mathbf{P}\left( \mathbf{\xi },\mathbf{\pi },0,\tau _{1}\right)
\right) \right\} _{\mathbf{\xi }=\mathbf{X}\left( \mathbf{x},\mathbf{p},\tau
_{1},\tau _{2}\right) ,\mathbf{\pi }=\mathbf{P}\left( \mathbf{x},\mathbf{p}%
,\tau _{1},\tau _{2}\right) -\hbar \mathbf{k}/2}  \label{34c}
\end{gather}

At time $\tau _{2}$ the $\pi $ pulse acts which, according to Eqs. (\ref{19}%
), transforms these density matrix elements at time $\tau _{2-}$ into 
\end{subequations}
\begin{subequations}
\label{35}
\begin{eqnarray}
\rho _{ee}\left( \mathbf{x},\mathbf{p},\tau _{2+}\right) &=&\frac{1}{2}%
f\left( \mathbf{X}\left( \mathbf{x},\mathbf{p}-\hbar \mathbf{k},0,\tau
_{2}\right) ,\mathbf{P}\left( \mathbf{x},\mathbf{p}-\hbar \mathbf{k},0,\tau
_{2}\right) \right) ,  \label{35a} \\
\rho _{gg}\left( \mathbf{x},\mathbf{p},\tau _{2+}\right) &=&\frac{1}{2}%
f\left( \mathbf{X}\left( \mathbf{\xi },\mathbf{\pi },0,\tau _{1}\right) ,%
\mathbf{P}\left( \mathbf{\xi },\mathbf{\pi },0,\tau _{1}\right) \right) _{%
\mathbf{\xi }=\mathbf{X}\left( \mathbf{x},\mathbf{p}+\hbar \mathbf{k},\tau
_{1},\tau _{2}\right) ,\mathbf{\pi }=\mathbf{P}\left( \mathbf{x},\mathbf{p}%
+\hbar \mathbf{k},\tau _{1},\tau _{2}\right) -\hbar \mathbf{k}}  \label{35b}
\\
\rho _{0eg}\left( \mathbf{x},\mathbf{p},\tau _{2+}\right) &=&\dfrac{i}{2}%
\left\{ \exp \left\{ i\left[ \mathbf{k\cdot }\left( 2\mathbf{x}-\mathbf{\xi }%
\right) -2\delta _{12}^{\left( 2\right) }\tau _{2}+\delta _{12}^{\left(
1\right) }\tau _{1}-2\phi _{2}+\phi _{1}\right] \right\} \right.  \notag \\
&&\left. \times f\left( \mathbf{X}\left( \mathbf{\xi },\mathbf{\pi },0,\tau
_{1}\right) ,\mathbf{P}\left( \mathbf{\xi },\mathbf{\pi },0,\tau _{1}\right)
\right) \right\} _{\mathbf{\xi }=\mathbf{X}\left( \mathbf{x},\mathbf{p},\tau
_{1},\tau _{2}\right) ,\mathbf{\pi }=\mathbf{P}\left( \mathbf{x},\mathbf{p}%
,\tau _{1},\tau _{2}\right) -\hbar \mathbf{k}/2},  \label{35c} \\
\rho _{Qeg}\left( \mathbf{x},\mathbf{p},\tau _{2+}\right) &=&-\dfrac{\hbar
^{2}}{48}\left\{ \exp \left\{ i\left[ \mathbf{k\cdot }\left( 2\mathbf{x}-%
\mathbf{\xi }\right) -2\delta _{12}^{\left( 2\right) }\tau _{2}+\delta
_{12}^{\left( 1\right) }\tau _{1}-2\phi _{2}+\phi _{1}\right] \right\}
\right.  \notag \\
&&\left. \times f\left( \mathbf{X}\left( \mathbf{\xi },\mathbf{\pi },0,\tau
_{1}\right) ,\mathbf{P}\left( \mathbf{\xi },\mathbf{\pi },0,\tau _{1}\right)
\right) \right\} _{\mathbf{\xi }=\mathbf{X}\left( \mathbf{x},\mathbf{p},\tau
_{1},\tau _{2}\right) ,\mathbf{\pi }=\mathbf{P}\left( \mathbf{x},\mathbf{p}%
,\tau _{1},\tau _{2}\right) -\hbar \mathbf{k}/2}  \notag \\
&&\times k_{u}k_{v}k_{w}\int_{\tau _{1}}^{\tau _{2}}dt\left[ \chi
_{ikl}^{\prime }\left( \mathbf{\xi },t\right) \right.  \notag \\
&&\times \left. \partial _{\mathbf{\pi }_{i}}\mathbf{X}_{u}\left( \mathbf{%
\xi },\mathbf{\pi },\tau _{1},t\right) \partial _{\mathbf{\pi }_{k}}\mathbf{X%
}_{v}\left( \mathbf{\xi },\mathbf{\pi },\tau _{1},t\right) \partial _{%
\mathbf{\pi }_{l}}\mathbf{X}_{w}\left( \mathbf{\xi },\mathbf{\pi },\tau
_{1},t\right) \right] _{\left\{ 
\begin{array}{c}
_{\mathbf{\xi }} \\ 
_{\mathbf{\pi }}%
\end{array}%
\right\} =\left\{ 
\begin{array}{c}
_{\mathbf{X}} \\ 
_{\mathbf{P}}%
\end{array}%
\right\} \left( \mathbf{x},\mathbf{p},t,\tau _{2}\right) }.  \label{35d}
\end{eqnarray}

The next step is to calculate the $Q-$term corrections in time interval $%
\left[ \tau _{2},\tau _{3}\right] .$ Each density matrix element in Eqs. (%
\ref{35}) produces a $Q-$term correction. However, the diagonal matrix
elements given by Eqs. (\ref{35a}, \ref{35b}) contain no rapidly oscillating
phase factors in momentum space allowing us to neglect their $Q-$term
corrections. Moreover, Eq. (\ref{35d}) is already linear in $Q$ and can
produce only higher order corrections that we neglect in this work. As a
consequence, we need consider only the $Q-$term correction produced by the
coherence in Eq. (\ref{35c}), which we denote as $\rho _{Qeg}^{\prime }.$
From Eq. (\ref{15}) we find 
\end{subequations}
\begin{eqnarray}
\rho _{Qeg}^{\prime }\left( \mathbf{x},\mathbf{p},\tau _{3-}\right) &=&i%
\QDABOVE{1pt}{\hbar ^{2}}{48}\int_{\tau _{2}}^{\tau _{3}}dt\left\{ \chi
_{ikl}^{\prime }\left( \mathbf{\xi },t\right) \partial _{\mathbf{\pi }%
_{i}}\partial _{\mathbf{\pi }_{k}}\partial _{\mathbf{\pi }_{l}}\right. 
\notag \\
&&\times \exp \left\{ i\left[ \mathbf{k}\cdot \left( 2\mathbf{X}\left( 
\mathbf{\xi },\mathbf{\pi },\tau _{2},t\right) -\mathbf{X}\left( \mathbf{\xi 
},\mathbf{\pi },\tau _{1},t\right) \right) \right. \right.  \notag \\
&&-\left. \left. 2\delta _{12}^{\left( 2\right) }\tau _{2}+\delta
_{12}^{\left( 1\right) }\tau _{1}-2\phi _{2}+\phi _{1}\right] \right\} 
\notag \\
&&\times f\left( \mathbf{X}\left( \mathbf{X}\left( \mathbf{\xi ,\pi },\tau
_{1},t\right) ,\mathbf{P}\left( \mathbf{\xi ,\pi },\tau _{1},t\right) -\frac{%
\hbar \mathbf{k}}{2},0,\tau _{1}\right) \right.  \notag \\
&&\left. \left. \mathbf{P}\left( \mathbf{X}\left( \mathbf{\xi ,\pi },\tau
_{1},t\right) ,\mathbf{P}\left( \mathbf{\xi ,\pi },\tau _{1},t\right) -\frac{%
\hbar \mathbf{k}}{2},0,\tau _{1}\right) \right) \right\} _{\left\{ 
\begin{array}{c}
_{\mathbf{\xi }} \\ 
_{\mathbf{\pi }}%
\end{array}%
\right\} =\left\{ 
\begin{array}{c}
_{\mathbf{X}} \\ 
_{\mathbf{P}}%
\end{array}%
\right\} \left( \mathbf{x},\mathbf{p},t,\tau _{3}\right) }  \label{36}
\end{eqnarray}%
where we used the multiplication law (\ref{31}), 
\begin{equation}
\left\{ 
\begin{array}{c}
\mathbf{X} \\ 
\mathbf{P}%
\end{array}%
\right\} \left( \mathbf{X}\left( \mathbf{\xi },\mathbf{\pi },\tau
_{2},t\right) ,\mathbf{P}\left( \mathbf{\xi },\mathbf{\pi },\tau
_{2},t\right) ,\tau _{1},\tau _{2}\right) =\left\{ 
\begin{array}{c}
\mathbf{X} \\ 
\mathbf{P}%
\end{array}%
\right\} \left( \mathbf{\xi },\mathbf{\pi },\tau _{1},t\right) .
\label{36.1}
\end{equation}

In Eq. (\ref{36}) the differentiation over momentum $\mathbf{\pi }$ is
carried out only for the Doppler phase factors. After differentiation, we
apply the multiplication law two more times to the phase factor and
distribution $f$, namely%
\begin{equation}
\left\{ 
\begin{tabular}{l}
$\mathbf{X}$ \\ 
$\mathbf{P}$%
\end{tabular}%
\right\} \left( \mathbf{\xi },\mathbf{\pi },\tau _{i},t\right) _{\left\{ 
\begin{array}{c}
_{\mathbf{\xi }} \\ 
_{\mathbf{\pi }}%
\end{array}%
\right\} =\left\{ 
\begin{array}{c}
_{\mathbf{X}} \\ 
_{\mathbf{P}}%
\end{array}%
\right\} \left( \mathbf{x},\mathbf{p},t,\tau _{3}\right) }=\left\{ 
\begin{tabular}{l}
$\mathbf{X}$ \\ 
$\mathbf{P}$%
\end{tabular}%
\right\} \left( \mathbf{x},\mathbf{p},\tau _{i},\tau _{3}\right)  \label{37}
\end{equation}%
for $i=1,2,$ and find that these terms become $t-$independent. As a result
one gets for $Q-$term $\rho _{Qeg}^{\prime }$ before the third pulse,%
\begin{eqnarray}
\rho _{Qeg}^{\prime }\left( \mathbf{x},\mathbf{p},\tau _{3-}\right) &=&-%
\QDABOVE{1pt}{\hbar ^{2}}{48}\left\{ \exp \left\{ i\left[ \mathbf{k}\cdot
\left( 2\mathbf{X}\left( \mathbf{x},\mathbf{p},\tau _{2},\tau _{3}\right) -%
\mathbf{\xi }\right) -2\delta _{12}^{\left( 2\right) }\tau _{2}+\delta
_{12}^{\left( 1\right) }\tau _{1}-2\phi _{2}+\phi _{1}\right] \right\}
\right.  \notag \\
&&\left. \times f\left( \mathbf{X}\left( \mathbf{\xi },\mathbf{\pi },0,\tau
_{1}\right) ,\mathbf{P}\left( \mathbf{\xi },\mathbf{\pi },0,\tau _{1}\right)
\right) \right\} _{\mathbf{\xi }=\mathbf{X}\left( \mathbf{x},\mathbf{p},\tau
_{1},\tau _{3}\right) ,\mathbf{\pi }=\mathbf{P}\left( \mathbf{x},\mathbf{p}%
,\tau _{1},\tau _{3}\right) -\hbar \mathbf{k}/2}  \notag \\
&&\times k_{u}k_{v}k_{w}\int_{\tau _{2}}^{\tau _{3}}dt\left\{ \chi
_{ikl}^{\prime }\left( \mathbf{\xi },t\right) \left[ \frac{\partial \mathbf{X%
}_{u}\left( \mathbf{\xi },\mathbf{\pi },\tau _{1},t\right) }{\partial 
\mathbf{\pi }_{i}}-2\frac{\partial \mathbf{X}_{u}\left( \mathbf{\xi },%
\mathbf{\pi },\tau _{2},t\right) }{\partial \mathbf{\pi }_{i}}\right] \right.
\notag \\
&&\times \left[ \frac{\partial \mathbf{X}_{v}\left( \mathbf{\xi },\mathbf{%
\pi },\tau _{1},t\right) }{\partial \mathbf{\pi }_{k}}-2\frac{\partial 
\mathbf{X}_{v}\left( \mathbf{\xi },\mathbf{\pi },\tau _{2},t\right) }{%
\partial \mathbf{\pi }_{k}}\right]  \notag \\
&&\times \left. \left[ \frac{\partial \mathbf{X}_{w}\left( \mathbf{\xi },%
\mathbf{\pi },\tau _{1},t\right) }{\partial \mathbf{\pi }_{l}}-2\frac{%
\partial \mathbf{X}_{w}\left( \mathbf{\xi },\mathbf{\pi },\tau _{2},t\right) 
}{\partial \mathbf{\pi }_{l}}\right] \right\} _{\left\{ 
\begin{array}{c}
_{\mathbf{\xi }} \\ 
_{\mathbf{\pi }}%
\end{array}%
\right\} =\left\{ 
\begin{array}{c}
_{\mathbf{X}} \\ 
_{\mathbf{P}}%
\end{array}%
\right\} \left( \mathbf{x},\mathbf{p},t,\tau _{3}\right) }.  \label{39}
\end{eqnarray}

The value of $\rho _{ee}\left( \mathbf{x},\mathbf{p},\tau _{3+}\right) $
will depend both on $\rho _{Qeg}^{\prime }\left( \mathbf{x},\mathbf{p},\tau
_{3-}\right) $ and $\rho _{0}\left( \mathbf{x},\mathbf{p},\tau _{3-}\right)
. $ In other words, we must also calculate the time evolution of $\rho
_{0}\left( \mathbf{x},\mathbf{p},t\right) $ between the second and third
pulses. Applying Eq. (\ref{12}), we find 
\begin{subequations}
\label{40}
\begin{eqnarray}
\rho _{ee}\left( \mathbf{x},\mathbf{p},\tau _{3-}\right) &=&\frac{1}{2}%
f\left( \mathbf{X}\left( \mathbf{\xi },\mathbf{\pi },0,\tau _{2}\right) ,%
\mathbf{P}\left( \mathbf{\xi },\mathbf{\pi },0,\tau _{2}\right) \right) _{%
\mathbf{\xi }=\mathbf{X}\left( \mathbf{x},\mathbf{p},\tau _{2},\tau
_{3}\right) ,\mathbf{\pi }=\mathbf{P}\left( \mathbf{x},\mathbf{p},\tau
_{2},\tau _{3}\right) -\hbar \mathbf{k}};  \label{40a} \\
\rho _{gg}\left( \mathbf{x},\mathbf{p},\tau _{3-}\right) &=&\frac{1}{2}%
f\left( \mathbf{X}\left( \mathbf{\xi },\mathbf{\pi },0,\tau _{1}\right) ,%
\mathbf{P}\left( \mathbf{\xi },\mathbf{\pi },0,\tau _{1}\right) \right) _{%
\begin{array}{l}
_{\left\{ \mathbf{\xi }=\mathbf{X}\left( \mathbf{X}\left( \mathbf{x},\mathbf{%
p},\tau _{2},\tau _{3}\right) ,\mathbf{P}\left( \mathbf{x},\mathbf{p},\tau
_{2},\tau _{3}\right) +\hbar \mathbf{k},\tau _{1},\tau _{2}\right) ,\right. }
\\ 
_{\left. \mathbf{\pi }=\mathbf{P}\left( \mathbf{X}\left( \mathbf{x},\mathbf{p%
},\tau _{2},\tau _{3}\right) ,\mathbf{P}\left( \mathbf{x},\mathbf{p},\tau
_{2},\tau _{3}\right) +\hbar \mathbf{k},\tau _{1},\tau _{2}\right) -\hbar 
\mathbf{k}\right\} }%
\end{array}%
}  \label{40b} \\
\rho _{0eg}\left( \mathbf{x},\mathbf{p},\tau _{3-}\right) &=&\dfrac{i}{2}%
\left\{ \exp \left\{ i\left[ \mathbf{k}\cdot \left( 2\mathbf{X}\left( 
\mathbf{x},\mathbf{p},\tau _{2},\tau _{3}\right) -\mathbf{\xi }\right)
-2\delta _{12}^{\left( 2\right) }\tau _{2}+\delta _{12}^{\left( 1\right)
}\tau _{1}-2\phi _{2}+\phi _{1}\right] \right\} \right.  \notag \\
&&\left. \times f\left( \mathbf{X}\left( \mathbf{\xi },\mathbf{\pi },0,\tau
_{1}\right) ,\mathbf{P}\left( \mathbf{\xi },\mathbf{\pi },0,\tau _{1}\right)
\right) \right\} _{\mathbf{\xi }=\mathbf{X}\left( \mathbf{x},\mathbf{p},\tau
_{1},\tau _{3}\right) ,\mathbf{\pi }=\mathbf{P}\left( \mathbf{x},\mathbf{p}%
,\tau _{1},\tau _{3}\right) -\hbar \mathbf{k}/2},  \label{40c} \\
\rho _{Qeg}\left( \mathbf{x},\mathbf{p},\tau _{3-}\right) &=&-\dfrac{\hbar
^{2}}{48}\left\{ \exp \left\{ i\left[ \mathbf{k}\cdot \left( 2\mathbf{X}%
\left( \mathbf{x},\mathbf{p},\tau _{2},\tau _{3}\right) -\mathbf{\xi }%
\right) -2\delta _{12}^{\left( 2\right) }\tau _{2}+\delta _{12}^{\left(
1\right) }\tau _{1}-2\phi _{2}+\phi _{1}\right] \right\} \right.  \notag \\
&&\left. \times f\left( \mathbf{X}\left( \mathbf{\xi },\mathbf{\pi },0,\tau
_{1}\right) ,\mathbf{P}\left( \mathbf{\xi },\mathbf{\pi },0,\tau _{1}\right)
\right) \right\} _{\mathbf{\xi }=\mathbf{X}\left( \mathbf{x},\mathbf{p},\tau
_{1},\tau _{3}\right) ,\mathbf{\pi }=\mathbf{P}\left( \mathbf{x},\mathbf{p}%
,\tau _{1},\tau _{3}\right) -\hbar \mathbf{k}/2}  \notag \\
&&\times \mathbf{k}_{u}\mathbf{k}_{v}\mathbf{k}_{w}\int_{\tau _{1}}^{\tau
_{2}}dt\left[ \chi _{ikl}^{\prime }\left( \mathbf{\xi },t\right) \partial _{%
\mathbf{\pi }_{i}}\mathbf{X}_{u}\left( \mathbf{\xi },\mathbf{\pi },\tau
_{1},t\right) \partial _{\mathbf{\pi }_{k}}\mathbf{X}_{v}\left( \mathbf{\xi }%
,\mathbf{\pi },\tau _{1},t\right) \right.  \notag \\
&&\times \left. \partial _{\mathbf{\pi }_{l}}\mathbf{X}_{w}\left( \mathbf{%
\xi },\mathbf{\pi },\tau _{1},t\right) \right] _{\left\{ 
\begin{array}{c}
_{\mathbf{\xi }} \\ 
_{\mathbf{\pi }}%
\end{array}%
\right\} =\left\{ 
\begin{array}{c}
_{\mathbf{X}} \\ 
_{\mathbf{P}}%
\end{array}%
\right\} \left( \mathbf{x},\mathbf{p},t,\tau _{3}\right) }.  \label{40d}
\end{eqnarray}

Combining the different contributions to the off diagonal density matrix
element given by Eqs. (\ref{40c}, \ref{40d}, \ref{39}) and factoring out a
common phase factor, we obtain 
\end{subequations}
\begin{subequations}
\label{41}
\begin{eqnarray}
\rho _{eg}\left( \mathbf{x},\mathbf{p},\tau _{3-}\right) &\approx &\rho
_{0eg}\left( \mathbf{x},\mathbf{p},\tau _{3-}\right) +\rho _{Qeg}\left( 
\mathbf{x},\mathbf{p},\tau _{3-}\right) +\rho _{Qeg}^{\prime }\left( \mathbf{%
x},\mathbf{p},\tau _{3-}\right)  \notag \\
&=&\dfrac{i}{2}\left\{ \exp \left\{ i\left[ \mathbf{k}\cdot \left( 2\mathbf{X%
}\left( \mathbf{x},\mathbf{p},\tau _{2},\tau _{3}\right) -\mathbf{\xi }%
\right) -2\delta _{12}^{\left( 2\right) }\tau _{2}+\delta _{12}^{\left(
1\right) }\tau _{1}-2\phi _{2}+\phi _{1}\right] \right\} \right.  \notag \\
&&\left. \times f\left( \mathbf{X}\left( \mathbf{\xi },\mathbf{\pi },0,\tau
_{1}\right) ,\mathbf{P}\left( \mathbf{\xi },\mathbf{\pi },0,\tau _{1}\right)
\right) \right\} _{\mathbf{\xi }=\mathbf{X}\left( \mathbf{x},\mathbf{p},\tau
_{1},\tau _{3}\right) ,\mathbf{\pi }=\mathbf{P}\left( \mathbf{x},\mathbf{p}%
,\tau _{1},\tau _{3}\right) -\hbar \mathbf{k}/2}\left[ 1-i\tilde{\phi}%
_{Q}\left( \mathbf{x},\mathbf{p}\right) \right] \\
&\approx &\dfrac{i}{2}\left\{ \exp \left\{ i\left[ \mathbf{k}\cdot \left( 2%
\mathbf{X}\left( \mathbf{x},\mathbf{p},\tau _{2},\tau _{3}\right) -\mathbf{%
\xi }\right) -\tilde{\phi}_{Q}\left( \mathbf{x},\mathbf{p}\right) -2\delta
_{12}^{\left( 2\right) }\tau _{2}+\delta _{12}^{\left( 1\right) }\tau
_{1}-2\phi _{2}+\phi _{1}\right] \right\} \right.  \notag \\
&&\left. \times f\left( \mathbf{X}\left( \mathbf{\xi },\mathbf{\pi },0,\tau
_{1}\right) ,\mathbf{P}\left( \mathbf{\xi },\mathbf{\pi },0,\tau _{1}\right)
\right) \right\} _{\mathbf{\xi }=\mathbf{X}\left( \mathbf{x},\mathbf{p},\tau
_{1},\tau _{3}\right) ,\mathbf{\pi }=\mathbf{P}\left( \mathbf{x},\mathbf{p}%
,\tau _{1},\tau _{3}\right) -\hbar \mathbf{k}/2},  \label{41a} \\
\tilde{\phi}_{Q}\left( \mathbf{x},\mathbf{p}\right) &=&-\dfrac{\hbar ^{2}}{24%
}k_{u}k_{v}k_{w}\left\{ \int_{\tau _{1}}^{\tau _{2}}dt\chi _{ikl}^{\prime
}\left( \mathbf{\xi }^{\prime },t\right) \partial _{\mathbf{\pi }%
_{i}^{\prime }}\mathbf{X}_{u}\left( \mathbf{\xi }^{\prime },\mathbf{\pi }%
^{\prime },\tau _{1},t\right) \partial _{\mathbf{\pi }_{k}^{\prime }}\mathbf{%
X}_{v}\left( \mathbf{\xi }^{\prime },\mathbf{\pi }^{\prime },\tau
_{1},t\right) \partial _{\mathbf{\pi }_{l}^{\prime }}\mathbf{X}_{w}\left( 
\mathbf{\xi }^{\prime },\mathbf{\pi }^{\prime },\tau _{1},t\right) \right. 
\notag \\
&&+\int_{\tau _{2}}^{\tau _{3}}dt\chi _{ikl}^{\prime }\left( \mathbf{\xi }%
^{\prime },t\right) \left[ \frac{\partial \mathbf{X}_{u}\left( \mathbf{\xi }%
^{\prime },\mathbf{\pi }^{\prime },\tau _{1},t\right) }{\partial \mathbf{\pi 
}_{i}^{\prime }}-2\frac{\partial \mathbf{X}_{u}\left( \mathbf{\xi }^{\prime
},\mathbf{\pi }^{\prime },\tau _{2},t\right) }{\partial \mathbf{\pi }%
_{i}^{\prime }}\right] \left[ \frac{\partial \mathbf{X}_{v}\left( \mathbf{%
\xi }^{\prime },\mathbf{\pi }^{\prime },\tau _{1},t\right) }{\partial 
\mathbf{\pi }_{k}^{\prime }}-2\frac{\partial \mathbf{X}_{v}\left( \mathbf{%
\xi }^{\prime },\mathbf{\pi }^{\prime },\tau _{2},t\right) }{\partial 
\mathbf{\pi }_{k}^{\prime }}\right]  \notag \\
&&\times \left. \left[ \frac{\partial \mathbf{X}_{w}\left( \mathbf{\xi }%
^{\prime },\mathbf{\pi }^{\prime },\tau _{1},t\right) }{\partial \mathbf{\pi 
}_{l}^{\prime }}-2\frac{\partial \mathbf{X}_{w}\left( \mathbf{\xi }^{\prime
},\mathbf{\pi }^{\prime },\tau _{2},t\right) }{\partial \mathbf{\pi }%
_{l}^{\prime }}\right] \right\} _{\left\{ 
\begin{array}{c}
_{\mathbf{\xi }^{\prime }} \\ 
_{\mathbf{\pi }^{\prime }}%
\end{array}%
\right\} =\left\{ 
\begin{array}{c}
_{\mathbf{X}} \\ 
_{\mathbf{P}}%
\end{array}%
\right\} \left( \mathbf{x},\mathbf{p},t,\tau _{3}\right) }.  \label{41b}
\end{eqnarray}

Finally we use Eqs. (\ref{18a}, \ref{40a}, \ref{40b}, \ref{41}) to calculate 
$\rho _{ee}\left( \mathbf{x},\mathbf{p},\tau _{3+}\right) $ following the $%
\pi /2$ pulse at time $\tau _{3}$ as 
\end{subequations}
\begin{eqnarray}
\rho _{ee}\left( \mathbf{x},\mathbf{p},\tau _{3+}\right) &=&\frac{1}{4}%
f\left( \mathbf{X}\left( \mathbf{\xi },\mathbf{\pi },0,\tau _{2}\right) ,%
\mathbf{P}\left( \mathbf{\xi },\mathbf{\pi },0,\tau _{2}\right) \right) _{%
\mathbf{\xi }=\mathbf{X}\left( \mathbf{x},\mathbf{p},\tau _{2},\tau
_{3}\right) ,\mathbf{\pi }=\mathbf{P}\left( \mathbf{x},\mathbf{p},\tau
_{2},\tau _{3}\right) -\hbar \mathbf{k}}  \notag \\
&&+\frac{1}{4}f\left( \mathbf{X}\left( \mathbf{\xi },\mathbf{\pi },0,\tau
_{1}\right) ,\mathbf{P}\left( \mathbf{\xi },\mathbf{\pi },0,\tau _{1}\right)
\right) _{\left\{ 
\begin{array}{c}
_{\mathbf{\xi }} \\ 
_{\mathbf{\pi }}%
\end{array}%
\right\} =\left\{ 
\begin{array}{l}
_{\mathbf{X}\left( \mathbf{X}\left( \mathbf{x},\mathbf{p}-\hbar \mathbf{k}%
,\tau _{2},\tau _{3}\right) ,\mathbf{P}\left( \mathbf{x},\mathbf{p}-\hbar 
\mathbf{k},\tau _{2},\tau _{3}\right) +\hbar \mathbf{k},\tau _{1},\tau
_{2}\right) } \\ 
_{\mathbf{P}\left( \mathbf{X}\left( \mathbf{x},\mathbf{p}-\hbar \mathbf{k}%
,\tau _{2},\tau _{3}\right) ,\mathbf{P}\left( \mathbf{x},\mathbf{p}-\hbar 
\mathbf{k},\tau _{2},\tau _{3}\right) +\hbar \mathbf{k},\tau _{1},\tau
_{2}\right) -\hbar \mathbf{k}}%
\end{array}%
\right\} }  \notag \\
&&-\frac{1}{2}\left\{ \cos \left[ \mathbf{k}\cdot \left( \mathbf{x}-2\mathbf{%
X}\left( \mathbf{x},\mathbf{p}-\frac{\hbar \mathbf{k}}{2},\tau _{2},\tau
_{3}\right) +\mathbf{\xi }\right) +\tilde{\phi}_{Q}\left( \mathbf{x},\mathbf{%
p}-\frac{\hbar \mathbf{k}}{2}\right) \right. \right.  \notag \\
&&\left. -\delta _{12}^{\left( 3\right) }\tau _{3}+2\delta _{12}^{\left(
2\right) }\tau _{2}-\delta _{12}^{\left( 1\right) }\tau _{1}-\phi _{3}+2\phi
_{2}-\phi _{1}\right]  \notag \\
&&\left. \times f\left( \mathbf{X}\left( \mathbf{\xi },\mathbf{\pi },0,\tau
_{1}\right) ,\mathbf{P}\left( \mathbf{\xi },\mathbf{\pi },0,\tau _{1}\right)
\right) \right\} _{\mathbf{\xi }=\mathbf{X}\left( \mathbf{x},\mathbf{p}%
-\hbar \mathbf{k}/2,\tau _{1},\tau _{3}\right) ,\mathbf{\pi }=\mathbf{P}%
\left( \mathbf{x},\mathbf{p}-\hbar \mathbf{k}/2,\tau _{1},\tau _{3}\right)
-\hbar \mathbf{k}/2}.  \label{42}
\end{eqnarray}

This density matrix element can be used to calculate any physically measured
observable associated with atoms in state $e$. For example, one could
measure the state $e$ population given as%
\begin{equation}
w=\int d\mathbf{x}d\mathbf{p}\rho _{ee}\left( \mathbf{x},\mathbf{p},\tau
_{3+}\right) .  \label{43}
\end{equation}%
The first two terms in Eq. (\ref{42}) are responsible for the background
signal. When substituted into Eq. (\ref{43}), they yield a background
contribution equal to $1/2$, allowing us to write 
\begin{equation}
w=\frac{1}{2}\left( 1-\tilde{w}\right) ,  \label{44}
\end{equation}%
where the interferometric term $\tilde{w}$ is given by%
\begin{eqnarray}
\tilde{w} &=&\int d\mathbf{x}d\mathbf{p}\left\{ \cos \left[ \mathbf{k}\cdot
\left( \mathbf{x}-2\mathbf{X}\left( \mathbf{x},\mathbf{p}-\frac{\hbar 
\mathbf{k}}{2},\tau _{2},\tau _{3}\right) +\mathbf{\xi }\right) +\tilde{\phi}%
_{Q}\left( \mathbf{x},\mathbf{p}-\frac{\hbar \mathbf{k}}{2}\right) -\delta
_{12}^{\left( 3\right) }\tau _{3}+2\delta _{12}^{\left( 2\right) }\tau
_{2}-\delta _{12}^{\left( 1\right) }\tau _{1}-\phi _{3}+2\phi _{2}-\phi _{1}%
\right] \right.  \notag \\
&&\left. \times f\left( \mathbf{X}\left( \mathbf{\xi },\mathbf{\pi },0,\tau
_{1}\right) ,\mathbf{P}\left( \mathbf{\xi },\mathbf{\pi },0,\tau _{1}\right)
\right) \right\} _{\mathbf{\xi }=\mathbf{X}\left( \mathbf{x},\mathbf{p}%
-\hbar \mathbf{k}/2,\tau _{1},\tau _{3}\right) ,\mathbf{\pi }=\mathbf{P}%
\left( \mathbf{x},\mathbf{p}-\hbar \mathbf{k}/2,\tau _{1},\tau _{3}\right)
-\hbar \mathbf{k}/2}.  \label{45}
\end{eqnarray}

To carry out the integration we express all position and momenta in terms of
the position and momentum variables at time $0$, denoted by 
\begin{equation}
\left\{ \mathbf{x}^{\prime },\mathbf{p}^{\prime }\right\} =\left\{ \mathbf{X}%
\left( \mathbf{\xi },\mathbf{\pi },0,\tau _{1}\right) ,\mathbf{P}\left( 
\mathbf{\xi },\mathbf{\pi },0,\tau _{1}\right) \right\}  \label{46}
\end{equation}%
In terms of these variables, 
\begin{subequations}
\label{47}
\begin{eqnarray}
\left\{ \mathbf{\xi },\mathbf{\pi }\right\} &=&\left\{ \mathbf{X}\left( 
\mathbf{x}^{\prime },\mathbf{p}^{\prime },\tau _{1},0\right) ,\mathbf{P}%
\left( \mathbf{x}^{\prime },\mathbf{p}^{\prime },\tau _{1},0\right) \right\}
,  \label{47a} \\
\left\{ \mathbf{x},\mathbf{p}\right\} &=&\left\{ \mathbf{X}\left( \mathbf{X}%
\left( \mathbf{x}^{\prime },\mathbf{p}^{\prime },\tau _{1},0\right) ,\mathbf{%
P}\left( \mathbf{x}^{\prime },\mathbf{p}^{\prime },\tau _{1},0\right) +\hbar 
\mathbf{k}/2,\tau _{3},\tau _{1}\right) ,\right.  \notag \\
&&\left. \mathbf{P}\left( \mathbf{X}\left( \mathbf{x}^{\prime },\mathbf{p}%
^{\prime },\tau _{1},0\right) ,\mathbf{P}\left( \mathbf{x}^{\prime },\mathbf{%
p}^{\prime },\tau _{1},0\right) +\hbar \mathbf{k}/2,\tau _{3},\tau
_{1}\right) +\hbar \mathbf{k}/2\right\} ,  \label{47b} \\
\left\vert \partial \left\{ \mathbf{x},\mathbf{p}\right\} /\partial \left\{ 
\mathbf{x}^{\prime },\mathbf{p}^{\prime }\right\} \right\vert &=&1,
\label{47c} \\
\mathbf{X}\left( \mathbf{x},\mathbf{p}-\hbar \mathbf{k/2},\tau _{2},\tau
_{3}\right) &=&\mathbf{X}\left( \mathbf{X}\left( \mathbf{x}^{\prime },%
\mathbf{p}^{\prime },\tau _{1},0\right) ,\mathbf{P}\left( \mathbf{x}^{\prime
},\mathbf{p}^{\prime },\tau _{1},0\right) +\hbar \mathbf{k}/2,\tau _{2},\tau
_{1}\right) .  \label{47d}
\end{eqnarray}%
After redefining $\left\{ \mathbf{x}^{\prime },\mathbf{p}^{\prime }\right\}
\rightarrow \left\{ \mathbf{x},\mathbf{p}\right\} ,$ one finds 
\end{subequations}
\begin{equation}
\tilde{w}=\int d\mathbf{x}d\mathbf{p}\cos \left[ \phi \left( \mathbf{x},%
\mathbf{p}\right) -\delta _{12}^{\left( 3\right) }\tau _{3}+2\delta
_{12}^{\left( 2\right) }\tau _{2}-\delta _{12}^{\left( 1\right) }\tau
_{1}-\phi _{3}+2\phi _{2}-\phi _{1}\right] f\left( \mathbf{x},\mathbf{p}%
\right) ,  \label{47.1}
\end{equation}%
where the phase $\phi \left( \mathbf{x},\mathbf{p}\right) $ of the AI is
defined as 
\begin{subequations}
\label{49}
\begin{eqnarray}
\phi \left( \mathbf{x},\mathbf{p}\right) &=&\phi _{r}\left( \mathbf{x},%
\mathbf{p}\right) +\phi _{Q}\left( \mathbf{x},\mathbf{p}\right) ;
\label{49a} \\
\phi _{r}\left( \mathbf{x},\mathbf{p}\right) &=&\mathbf{k}\cdot \left[ 
\mathbf{X}\left( \mathbf{\xi },\mathbf{\pi },\tau _{3},\tau _{1}\right) -2%
\mathbf{X}\left( \mathbf{\xi },\mathbf{\pi },\tau _{2},\tau _{1}\right) +%
\mathbf{\xi }\right] _{\left\{ \mathbf{\xi }=\mathbf{X}\left( \mathbf{x},%
\mathbf{p},\tau _{1},0\right) ,\mathbf{\pi }=\mathbf{P}\left( \mathbf{x},%
\mathbf{p},\tau _{1},0\right) +\hbar \mathbf{k}/2\right\} };  \label{49b} \\
\phi _{Q}\left( \mathbf{x},\mathbf{p}\right) &=&\tilde{\phi}_{Q}\left[ 
\mathbf{X}\left( \mathbf{\xi },\mathbf{\pi },\tau _{3},\tau _{1}\right) ,%
\mathbf{P}\left( \mathbf{\xi },\mathbf{\pi },\tau _{3},\tau _{1}\right) %
\right] _{\left\{ \mathbf{\xi }=\mathbf{X}\left( \mathbf{x},\mathbf{p},\tau
_{1},0\right) ,\mathbf{\pi }=\mathbf{P}\left( \mathbf{x},\mathbf{p},\tau
_{1},0\right) +\hbar \mathbf{k}/2\right\} },  \label{49c}
\end{eqnarray}%
with $\tilde{\phi}_{Q}$ given by Eq. (\ref{41b}).

\subsubsection{Atom trajectories in the presence of the test mass}

To calculate the phases in Eqs. (\ref{49}) we need expressions for the
propagation functions $\left\{ \mathbf{X}\left( \mathbf{x},\mathbf{p}%
,t,t^{\prime }\right) ,\mathbf{P}\left( \mathbf{x},\mathbf{p},t,t^{\prime
}\right) \right\} $, i.e. atomic position and momentum at time $t$ subject
to the initial value $\left\{ \mathbf{x},\mathbf{p}\right\} $ at time $%
t^{\prime }$. These functions evolve as 
\end{subequations}
\begin{subequations}
\label{100}
\begin{eqnarray}
\mathbf{\dot{X}}\left( \mathbf{x},\mathbf{p},t,t^{\prime }\right) &=&\dfrac{%
\mathbf{P}\left( \mathbf{x},\mathbf{p},t,t^{\prime }\right) }{M_{a}},
\label{100a} \\
\mathbf{\dot{P}}\left( \mathbf{x},\mathbf{p},t,t^{\prime }\right)
&=&M_{a}\left\{ \mathbf{g}+\delta \mathbf{g}\left[ \mathbf{X}\left( \mathbf{x%
},\mathbf{p},t,t^{\prime }\right) ,t\right] \right\} .  \label{100b}
\end{eqnarray}%
We neglect in Eq. (\ref{100}) the gravity-gradient, centrifugal and Coriolis
forces caused by the rotating Earth. When $\delta \mathbf{g}\left( \mathbf{x}%
,t\right) $ is a perturbation, the approximate solutions of Eqs. (\ref{100})
are \cite{a24} 
\end{subequations}
\begin{subequations}
\label{101}
\begin{eqnarray}
\mathbf{X}\left( \mathbf{x},\mathbf{p},t,t^{\prime }\right) &\approx &%
\mathbf{X}^{\left( 0\right) }\left( \mathbf{x},\mathbf{p},t,t^{\prime
}\right) +\delta \mathbf{X}\left( \mathbf{x},\mathbf{p},t,t^{\prime }\right)
,  \label{101a} \\
\mathbf{P}\left( \mathbf{x},\mathbf{p},t,t^{\prime }\right) &\approx &%
\mathbf{P}^{\left( 0\right) }\left( \mathbf{x},\mathbf{p},t,t^{\prime
}\right) +\delta \mathbf{P}\left( \mathbf{x},\mathbf{p},t,t^{\prime }\right)
,  \label{101b} \\
\mathbf{X}^{\left( 0\right) }\left( \mathbf{x},\mathbf{p},t,t^{\prime
}\right) &=&\mathbf{x}+\dfrac{\mathbf{p}}{M_{a}}\left( t-t^{\prime }\right) +%
\mathbf{g}\dfrac{\left( t-t^{\prime }\right) ^{2}}{2},  \label{101c} \\
\mathbf{P}^{\left( 0\right) }\left( \mathbf{x},\mathbf{p},t,t^{\prime
}\right) &=&\mathbf{p}+M_{a}\mathbf{g}\left( t-t^{\prime }\right) ;
\label{101d} \\
\delta \mathbf{X}\left( \mathbf{x},\mathbf{p},t,t^{\prime }\right)
&=&\int_{t^{\prime }}^{t}dt^{\prime \prime }\left( t-t^{\prime \prime
}\right) \delta \mathbf{g}\left[ \mathbf{X}^{\left( 0\right) }\left( \mathbf{%
x},\mathbf{p},t^{\prime \prime },t^{\prime }\right) ,t^{\prime \prime }%
\right]  \label{101e} \\
\delta \mathbf{P}\left( \mathbf{x},\mathbf{p},t,t^{\prime }\right)
&=&M_{a}\int_{t^{\prime }}^{t}dt^{\prime \prime }\delta \mathbf{g}\left[ 
\mathbf{X}^{\left( 0\right) }\left( \mathbf{x},\mathbf{p},t^{\prime \prime
},t^{\prime }\right) ,t^{\prime \prime }\right] .  \label{101f}
\end{eqnarray}%
Each of the functions $\left\{ \mathbf{X}^{\left( 0\right) },\mathbf{P}%
^{\left( 0\right) },\delta \mathbf{X},\delta \mathbf{P}\right\} $ obeys the
multiplication law (\ref{31}) 
\end{subequations}
\begin{subequations}
\label{102}
\begin{eqnarray}
\left\{ 
\begin{array}{c}
\mathbf{X}^{\left( 0\right) } \\ 
\mathbf{P}^{\left( 0\right) }%
\end{array}%
\right\} \left( \mathbf{X}^{\left( 0\right) }\left( \mathbf{x},\mathbf{p}%
,t^{\prime },t^{\prime \prime }\right) ,\mathbf{P}^{\left( 0\right) }\left( 
\mathbf{x},\mathbf{p},t^{\prime },t^{\prime \prime }\right) ,t,t^{\prime
}\right) &=&\left\{ 
\begin{array}{c}
\mathbf{X}^{\left( 0\right) } \\ 
\mathbf{P}^{\left( 0\right) }%
\end{array}%
\right\} \left( \mathbf{x},\mathbf{p},t,t^{\prime \prime }\right) ;
\label{102a} \\
\left\{ 
\begin{array}{c}
\delta \mathbf{X} \\ 
\delta \mathbf{P}%
\end{array}%
\right\} \left( \delta \mathbf{X}\left( \mathbf{x},\mathbf{p},t^{\prime
},t^{\prime \prime }\right) ,\delta \mathbf{P}\left( \mathbf{x},\mathbf{p}%
,t^{\prime },t^{\prime \prime }\right) ,t,t^{\prime }\right) &=&\left\{ 
\begin{array}{c}
\delta \mathbf{X} \\ 
\delta \mathbf{P}%
\end{array}%
\right\} \left( \mathbf{x},\mathbf{p},t,t^{\prime \prime }\right) .
\label{102b}
\end{eqnarray}

\subsubsection{Phases}

It remains for us to calculate the phases $\phi _{r}\left( \mathbf{x},%
\mathbf{p}\right) $ and $\phi _{Q}\left( \mathbf{x},\mathbf{p}\right) $. In
the following two subsections, we obtain both exact integral and approximate
integral and analytic expressions for these phases. In Sec. III, the exact
expressions are evaluated numerically and the range of validity of the
approximate expressions is established.

\paragraph{$\protect\phi _{r}\left( \mathbf{x},\mathbf{p}\right) $}

The phase $\phi _{r}$ includes a "classical" part (vanishing in the limit $%
\hbar \rightarrow 0),$ as well as a quantum correction $\phi _{q}$. The
contributions to $\phi _{r}$ resulting from the Earth's gravitational field
and the rotation of the Earth were calculated approximately in \cite{a14,c2}%
. The classical component of these contributions to $\phi _{r}$ has been
calculated exactly \cite{a36}. In this paper we concentrate on the additions
to $\phi _{r}$ caused by the test mass' field. The "classical" part of this
addition has been evaluated in Ref. \cite{a21}. Contributions to the phase
from the Earth's rotation and Earth's gravity-gradient terms are neglected
in this paper.

It is shown in the appendix how approximate expressions for the propagators
needed in Eq. (\ref{49b}) can be obtained from Eqs. (\ref{101}). It then
follows that the phase $\phi _{r}$ given in Eq. (\ref{49b}) can be written
as a sum of three terms, 
\end{subequations}
\begin{subequations}
\label{201}
\begin{eqnarray}
\phi _{r}\left( \mathbf{x},\mathbf{p}\right) &=&\phi _{0}\left( \mathbf{x},%
\mathbf{p}\right) +\delta \phi \left( \mathbf{x},\mathbf{p}\right) +\phi
_{q}\left( \mathbf{x},\mathbf{p}\right) ;  \label{201a} \\
\phi _{0}\left( \mathbf{x},\mathbf{p}\right) &=&\mathbf{k\cdot }\left[ 
\mathbf{X}^{\left( 0\right) }\left( \mathbf{x},\mathbf{p},\tau _{3},0\right)
-2\mathbf{X}^{\left( 0\right) }\left( \mathbf{x},\mathbf{p},\tau
_{2},0\right) +\mathbf{X}^{\left( 0\right) }\left( \mathbf{x},\mathbf{p}%
,\tau _{1},0\right) \right] =\mathbf{k\cdot g}T^{2},  \label{201b} \\
\delta \phi \left( \mathbf{x},\mathbf{p}\right) &=&\mathbf{k}\cdot \mathbf{%
\psi }\equiv \mathbf{k}\cdot \left[ \delta \mathbf{X}\left( \mathbf{x},%
\mathbf{p},\tau _{3},0\right) -2\delta \mathbf{X}\left( \mathbf{x},\mathbf{p}%
,\tau _{2},0\right) +\delta \mathbf{X}\left( \mathbf{x},\mathbf{p},\tau
_{1},0\right) \right] ,  \label{201d} \\
\phi _{q}\left( \mathbf{x},\mathbf{p}\right) &=&\mathbf{k}\cdot \mathbf{\psi 
}_{q},  \label{201e} \\
\mathbf{\psi }_{q} &=&\int_{\tau _{1}}^{\tau _{3}}dt\left( \tau
_{3}-t\right) \left\{ \delta \mathbf{g}\left[ \mathbf{X}^{\left( 0\right)
}\left( \mathbf{x},\mathbf{p},t,0\right) +\dfrac{\hbar \mathbf{k}}{2M_{a}}%
\left( t-\tau _{1}\right) ,t\right] -\delta \mathbf{g}\left[ \mathbf{X}%
^{\left( 0\right) }\left( \mathbf{x},\mathbf{p},t,0\right) ,t\right] \right\}
\notag \\
&&-2\int_{\tau _{1}}^{\tau _{2}}dt\left( \tau _{2}-t\right) \left\{ \delta 
\mathbf{g}\left[ \mathbf{X}^{\left( 0\right) }\left( \mathbf{x},\mathbf{p}%
,t,0\right) +\dfrac{\hbar \mathbf{k}}{2M_{a}}\left( t-\tau _{1}\right) ,t%
\right] -\delta \mathbf{g}\left[ \mathbf{X}^{\left( 0\right) }\left( \mathbf{%
x},\mathbf{p},t,0\right) ,t\right] \right\} .  \label{201f}
\end{eqnarray}%
The term $\phi _{0}\left( \mathbf{x},\mathbf{p}\right) $ is the classical
contribution from the Earth's field, the term $\delta \phi \left( \mathbf{x},%
\mathbf{p}\right) $ is the classical contribution from the test mass' field,
and the term $\phi _{q}\left( \mathbf{x},\mathbf{p}\right) $ is the quantum
correction to the test mass' field.

To evaluate the classical contribution to the phase given by Eq. (\ref{201d}%
), we use Eq. (\ref{101e}) to arrive at 
\end{subequations}
\begin{subequations}
\label{202}
\begin{align}
\mathbf{\psi }& =\tau _{3}\mathbf{u}_{20}-\tau _{1}\mathbf{u}_{10}+\mathbf{u}%
_{11}-\mathbf{u}_{21},  \label{202a} \\
\mathbf{u}_{\alpha \beta }& =\int_{\tau _{\alpha }}^{\tau _{\alpha
}+T}dt^{\prime \prime }\left( t^{\prime \prime }\right) ^{\beta }\delta 
\mathbf{g}\left[ \mathbf{X}^{\left( 0\right) }\left( \mathbf{x},\mathbf{p}%
,t^{\prime \prime },0\right) ,t^{\prime \prime }\right] .  \label{202b}
\end{align}%
Equations (\ref{201d}, \ref{202}) have been used in Ref. \cite{a21}. With
the simple change of variables, $t=\tau _{2}+\theta $ for $\mathbf{u}%
_{2\beta }$ and $t=\tau _{1}+\theta $ for $\mathbf{u}_{1\beta },$ we find 
\end{subequations}
\begin{equation}
\mathbf{\psi =}\int_{0}^{T}d\theta \left\{ \left( T-\theta \right) \delta 
\mathbf{g}\left[ \mathbf{X}^{\left( 0\right) }\left( \mathbf{x},\mathbf{p}%
,\tau _{2}+\theta ,0\right) ,\tau _{2}+\theta \right] +\theta \delta \mathbf{%
g}\left[ \mathbf{X}^{\left( 0\right) }\left( \mathbf{x},\mathbf{p},\tau
_{1}+\theta ,0\right) ,\tau _{1}+\theta \right] \right\} .  \label{202.1}
\end{equation}%
If the test mass moves without rotation and follows a trajectory denoted by $%
\mathbf{x}_{m}\left( t\right) $, then 
\begin{equation}
\delta \mathbf{g}\left( \mathbf{x,}t\right) =\delta \mathbf{g}\left[ \mathbf{%
x-x}_{m}\left( t\right) \right]  \label{202.2}
\end{equation}%
and%
\begin{equation}
\mathbf{\psi =}\int_{0}^{T}d\theta \left\{ \left( T-\theta \right) \delta 
\mathbf{g}\left[ \mathbf{X}^{\left( 0\right) }\left( \mathbf{x},\mathbf{p}%
,\tau _{2}+\theta ,0\right) \mathbf{-x}_{m}\left( \tau _{2}+\theta \right) %
\right] +\theta \delta \mathbf{g}\left[ \mathbf{X}^{\left( 0\right) }\left( 
\mathbf{x},\mathbf{p},\tau _{1}+\theta ,0\right) \mathbf{-x}_{m}\left( \tau
_{1}+\theta \right) \right] \right\} .  \label{202.3}
\end{equation}%
This is the exact expression for $\mathbf{\psi }$ that is used in Sec. III.

We can arrive at an approximate expression for $\mathbf{\psi }$ if we assume
that the distance between the atoms and the test mass is sufficiently large
to keep only those terms that are linear in the field gradient. In other
words, we can evaluate the field of the test mass at some average
displacement $\mathbf{x}_{C}$ between the test mass and the atoms'
trajectory. If we choose \cite{a34} 
\begin{equation}
\mathbf{x}_{C}=\dfrac{1}{T^{2}}\int_{0}^{T}d\theta \left\{ \left( T-\theta
\right) \left[ \mathbf{X}^{\left( 0\right) }\left( \mathbf{x},\mathbf{p}%
,\tau _{2}+\theta ,0\right) -\mathbf{x}_{m}\left( \tau _{2}+\theta \right) %
\right] +\theta \left[ \mathbf{X}^{\left( 0\right) }\left( \mathbf{x},%
\mathbf{p},\tau _{1}+\theta ,0\right) -\mathbf{x}_{m}\left( \tau _{1}+\theta
\right) \right] \right\} ,  \label{202.6}
\end{equation}%
expand 
\begin{equation}
\delta \mathbf{g}_{i}\left( \mathbf{x}\right) \approx \delta \mathbf{g}%
_{i}\left( \mathbf{x}_{C}\right) +\underline{\mathbf{\gamma }}\left( \mathbf{%
x}\right) \left( \mathbf{x}-\mathbf{x}_{C}\right) ,  \label{202.4}
\end{equation}%
where $\underline{\mathbf{\gamma }}\left( \mathbf{x}\right) $ is the
gravity-gradient tensor having matrix elements%
\begin{equation}
\underline{\mathbf{\gamma }}_{ij}\left( \mathbf{x}\right) =\frac{\partial
\delta \mathbf{g}_{i}\left( \mathbf{x}\right) }{\partial x_{j}},
\label{202.5}
\end{equation}%
and substitute the result back into Eq. (\ref{202.3}), we find that the term
proportional to $\underline{\mathbf{\gamma }}\left( \mathbf{x}\right) $
vanishes ($\mathbf{x}_{C}$ was chosen to insure this). We then obtain an
approximate expression $\delta \phi _{a}$ for the classical contribution to
the phase $\phi _{r}$ given by 
\begin{subequations}
\label{202.7}
\begin{eqnarray}
\delta \phi &\approx &\delta \phi _{a}=\mathbf{k\cdot \psi }_{a},
\label{202.7a} \\
\mathbf{\psi }_{a} &=&\delta \mathbf{g}\left( \mathbf{x}_{C}\right) T^{2}.
\label{202.7b}
\end{eqnarray}

We now turn our attention to the quantum correction. The vector $\mathbf{%
\psi }_{q}$ given in Eq. (\ref{201f}) can be rewritten as 
\end{subequations}
\begin{eqnarray}
\mathbf{\psi }_{q} &=&\int_{\tau _{2}}^{\tau _{3}}dt\left( \tau
_{3}-t\right) \left\{ \delta \mathbf{g}\left[ \mathbf{X}^{\left( 0\right)
}\left( \mathbf{x},\mathbf{p},t,0\right) +\dfrac{\hbar \mathbf{k}}{2M_{a}}%
\left( t-\tau _{1}\right) ,t\right] -\delta \mathbf{g}\left[ \mathbf{X}%
^{\left( 0\right) }\left( \mathbf{x},\mathbf{p},t,0\right) ,t\right] \right\}
\notag \\
&&+\int_{\tau _{1}}^{\tau _{2}}dt\left( t-\tau _{1}\right) \left\{ \delta 
\mathbf{g}\left[ \mathbf{X}^{\left( 0\right) }\left( \mathbf{x},\mathbf{p}%
,t,0\right) +\dfrac{\hbar \mathbf{k}}{2M_{a}}\left( t-\tau _{1}\right) ,t%
\right] -\delta \mathbf{g}\left[ \mathbf{X}^{\left( 0\right) }\left( \mathbf{%
x},\mathbf{p},t,0\right) ,t\right] \right\} .  \label{203}
\end{eqnarray}%
Substituting $t=\tau _{2}+\theta $ in the first term of Eq. (\ref{203}) and $%
t=\tau _{1}+\theta $ in the second term, we obtain%
\begin{eqnarray}
\mathbf{\psi }_{q} &=&\int_{0}^{T}d\theta \left\{ \left( T-\theta \right) %
\left[ \delta \mathbf{g}\left( \mathbf{X}^{\left( 0\right) }\left( \mathbf{x}%
,\mathbf{p},\tau _{2}+\theta ,0\right) +\dfrac{\hbar \mathbf{k}}{2M_{a}}%
\left( T+\theta \right) ,\tau _{2}+\theta \right) -\delta \mathbf{g}\left( 
\mathbf{X}^{\left( 0\right) }\left( \mathbf{x},\mathbf{p},\tau _{2}+\theta
,0\right) ,\tau _{2}+\theta \right) \right] \right.  \notag \\
&&+\left. \theta \left[ \delta \mathbf{g}\left( \mathbf{X}^{\left( 0\right)
}\left( \mathbf{x},\mathbf{p},\tau _{1}+\theta ,0\right) +\dfrac{\hbar 
\mathbf{k}}{2M_{a}}\theta ,\tau _{1}+\theta \right) -\delta \mathbf{g}\left( 
\mathbf{X}^{\left( 0\right) }\left( \mathbf{x},\mathbf{p},\tau _{1}+\theta
,0\right) ,\tau _{1}+\theta \right) \right] \right\} .  \label{204}
\end{eqnarray}%
For translational motion, when the gravitational field of the test mass is
given by Eq. (\ref{202.2}),%
\begin{eqnarray}
\mathbf{\psi }_{q} &=&\int_{0}^{T}d\theta \left\{ \left( T-\theta \right) %
\left[ \delta \mathbf{g}\left( \mathbf{X}^{\left( 0\right) }\left( \mathbf{x}%
,\mathbf{p},\tau _{2}+\theta ,0\right) +\dfrac{\hbar \mathbf{k}}{2M_{a}}%
\left( T+\theta \right) \mathbf{-x}_{m}\left( \tau _{2}+\theta \right)
\right) -\delta \mathbf{g}\left( \mathbf{X}^{\left( 0\right) }\left( \mathbf{%
x},\mathbf{p},\tau _{2}+\theta ,0\right) \mathbf{-x}_{m}\left( \tau
_{2}+\theta \right) \right) \right] \right.  \notag \\
&&+\left. \theta \left[ \delta \mathbf{g}\left( \mathbf{X}^{\left( 0\right)
}\left( \mathbf{x},\mathbf{p},\tau _{1}+\theta ,0\right) +\dfrac{\hbar 
\mathbf{k}}{2M_{a}}\theta \mathbf{-x}_{m}\left( \tau _{1}+\theta \right)
\right) -\delta \mathbf{g}\left( \mathbf{X}^{\left( 0\right) }\left( \mathbf{%
x},\mathbf{p},\tau _{1}+\theta ,0\right) \mathbf{-x}_{m}\left( \tau
_{1}+\theta \right) \right) \right] \right\} ,  \label{207.1aaa}
\end{eqnarray}%
This is the exact expression for $\mathbf{\psi }_{q}$ that is used in Sec.
III.

There are two approximate expressions we will derive for $\mathbf{\psi }$.
When the recoil effect is small,%
\begin{equation}
\dfrac{\hbar k}{2M_{a}}T\ll X^{\left( 0\right) }\left( \mathbf{x},\mathbf{p}%
,T,0\right) ,  \label{205}
\end{equation}%
we can expand the arguments in Eq. (\ref{207.1aaa}) to obtain a first
approximation $\phi _{q}\approx \phi _{qn}$ given by 
\begin{subequations}
\begin{eqnarray}
\phi _{q} &\approx &\phi _{qn}=\mathbf{k}\cdot \mathbf{\psi }_{qn}
\label{206a} \\
\mathbf{\psi }_{qn} &=&\int_{0}^{T}d\theta \left\{ \left( T^{2}-\theta
^{2}\right) \underline{\mathbf{\gamma }}\left[ \mathbf{X}^{\left( 0\right)
}\left( \mathbf{x},\mathbf{p},\tau _{2}+\theta ,0\right) ,\tau _{2}+\theta %
\right] +\theta ^{2}\underline{\mathbf{\gamma }}\left[ \mathbf{X}^{\left(
0\right) }\left( \mathbf{x},\mathbf{p},\tau _{1}+\theta ,0\right) ,\tau
_{1}+\theta \right] \right\} \dfrac{\hbar \mathbf{k}}{2M_{a}},  \label{206b}
\end{eqnarray}%
where 
\end{subequations}
\begin{equation}
\underline{\mathbf{\gamma }}_{ij}\left( \mathbf{x},t\right) =\frac{\partial 
\mathbf{g}_{i}\left( \mathbf{x},t\right) }{\partial x_{j}}.  \label{207}
\end{equation}%
For translational motion of the test mass, $\underline{\mathbf{\gamma }}%
\left( \mathbf{x},t\right) =\underline{\mathbf{\gamma }}\left[ \mathbf{x-x}%
_{m}\left( t\right) \right] $ and Eq. (\ref{206b}) reduces to 
\begin{equation}
\mathbf{\psi }_{qn}=\int_{0}^{T}d\theta \left\{ \left( T^{2}-\theta
^{2}\right) \underline{\mathbf{\gamma }}\left[ \mathbf{X}^{\left( 0\right)
}\left( \mathbf{x},\mathbf{p},\tau _{2}+\theta ,0\right) \mathbf{-x}%
_{m}\left( \tau _{2}+\theta \right) \right] +\theta ^{2}\underline{\mathbf{%
\gamma }}\left[ \mathbf{X}^{\left( 0\right) }\left( \mathbf{x},\mathbf{p}%
,\tau _{1}+\theta ,0\right) \mathbf{-x}_{m}\left( \tau _{1}+\theta \right) %
\right] \right\} \dfrac{\hbar \mathbf{k}}{2M_{a}}.  \label{207.1bbb}
\end{equation}

The second approximate expression we obtain for $\mathbf{\psi }_{q}$ is the
limit of Eq. (\ref{207.1bbb}) when the distance between the atoms and the
test mass is sufficiently large to keep only those terms that are linear in
the field gradient. If we choose%
\begin{equation}
\mathbf{x}_{qC}=\dfrac{1}{T^{3}}\int_{0}^{T}d\theta \left\{ \left(
T^{2}-\theta ^{2}\right) \left[ \mathbf{X}^{\left( 0\right) }\left( \mathbf{x%
},\mathbf{p},\tau _{2}+\theta ,0\right) -\mathbf{x}_{m}\left( \tau
_{2}+\theta \right) \right] +\theta ^{2}\left[ \mathbf{X}^{\left( 0\right)
}\left( \mathbf{x},\mathbf{p},\tau _{1}+\theta ,0\right) -\mathbf{x}%
_{m}\left( \tau _{1}+\theta \right) \right] \right\} ,  \label{210}
\end{equation}%
and expand 
\begin{equation}
\underline{\mathbf{\gamma }}_{ij}\left( \mathbf{x}\right) \approx \underline{%
\mathbf{\gamma }}_{ij}\left( \mathbf{x}_{qC}\right) +\chi _{ijl}\left( 
\mathbf{x}_{qC}\right) \left( \mathbf{x-x}_{qC}\right) _{l},  \label{208}
\end{equation}%
where%
\begin{equation}
\chi _{ijl}\left( \mathbf{x}\right) =\partial _{\mathbf{x}_{l}}\underline{%
\mathbf{\gamma }}_{ij}\left( \mathbf{x}\right)  \label{209}
\end{equation}%
is an element of the gravity curvature tensor, then the contribution from
the second term in Eq. (\ref{208}) vanishes and we find an approximate
expression $\phi _{qa}$ for the phase given by 
\begin{subequations}
\label{211}
\begin{eqnarray}
\phi _{q}\left( \mathbf{x},\mathbf{p}\right) &\approx &\phi _{qa}\left( 
\mathbf{x},\mathbf{p}\right) =\mathbf{k\cdot \psi }_{qa},  \label{211a} \\
\mathbf{\psi }_{qa} &=&\underline{\mathbf{\gamma }}\left( \mathbf{x}%
_{qC}\right) \dfrac{\hbar \mathbf{k}}{2M_{a}}T^{3}.  \label{211b}
\end{eqnarray}

\paragraph{$\protect\phi _{Q}\left( \mathbf{x},\mathbf{p}\right) $}

We now consider $Q-$term quantum corrections to the phase given by Eqs. (\ref%
{49c}, \ref{41b}). We first replace $\left\{ \mathbf{x},\mathbf{p}\right\} \ 
$by $\left\{ \mathbf{X}\left( \mathbf{\xi },\mathbf{\pi },\tau _{3},\tau
_{1}\right) ,\mathbf{P}\left( \mathbf{\xi },\mathbf{\pi },\tau _{3},\tau
_{1}\right) \right\} $ in Eq. (\ref{41b}) to obtain 
\end{subequations}
\begin{equation}
\left\{ 
\begin{array}{c}
_{\mathbf{\xi }^{\prime }} \\ 
_{\mathbf{\pi }^{\prime }}%
\end{array}%
\right\} =\left\{ 
\begin{array}{c}
_{\mathbf{X}} \\ 
_{\mathbf{P}}%
\end{array}%
\right\} \left( \mathbf{X}\left( \mathbf{\xi },\mathbf{\pi },\tau _{3},\tau
_{1}\right) ,\mathbf{P}\left( \mathbf{\xi },\mathbf{\pi },\tau _{3},\tau
_{1}\right) ,t,\tau _{3}\right) =\left\{ 
\begin{array}{c}
_{\mathbf{X}} \\ 
_{\mathbf{P}}%
\end{array}%
\right\} \left( \mathbf{\xi },\mathbf{\pi },t,\tau _{1}\right) ,
\label{49.1}
\end{equation}%
allowing us to write $\phi _{Q}\left( \mathbf{x},\mathbf{p}\right) $ as 
\begin{eqnarray}
\phi _{Q}\left( \mathbf{x},\mathbf{p}\right) &=&-\dfrac{\hbar ^{2}}{24}%
k_{u}k_{v}k_{w}\left\{ \int_{\tau _{1}}^{\tau _{2}}dt\chi _{ikl}^{\prime
}\left( \mathbf{\xi },t\right) \partial _{\mathbf{\pi }_{i}}\mathbf{X}%
_{u}\left( \mathbf{\xi },\mathbf{\pi },\tau _{1},t\right) \partial _{\mathbf{%
\pi }_{k}}\mathbf{X}_{v}\left( \mathbf{\xi },\mathbf{\pi },\tau
_{1},t\right) \partial _{\mathbf{\pi }_{l}}\mathbf{X}_{w}\left( \mathbf{\xi }%
,\mathbf{\pi },\tau _{1},t\right) \right.  \notag \\
&&+\int_{\tau _{2}}^{\tau _{3}}dt\chi _{ikl}^{\prime }\left( \mathbf{\xi }%
,t\right) \left[ \frac{\partial \mathbf{X}_{u}\left( \mathbf{\xi },\mathbf{%
\pi },\tau _{1},t\right) }{\partial \mathbf{\pi }_{i}}-2\frac{\partial 
\mathbf{X}_{u}\left( \mathbf{\xi },\mathbf{\pi },\tau _{2},t\right) }{%
\partial \mathbf{\pi }_{i}}\right] \left[ \frac{\partial \mathbf{X}%
_{v}\left( \mathbf{\xi },\mathbf{\pi },\tau _{1},t\right) }{\partial \mathbf{%
\pi }_{k}}-2\frac{\partial \mathbf{X}_{v}\left( \mathbf{\xi },\mathbf{\pi }%
,\tau _{2},t\right) }{\partial \mathbf{\pi }_{k}}\right]  \notag \\
&&\times \left. \left[ \frac{\partial \mathbf{X}_{w}\left( \mathbf{\xi },%
\mathbf{\pi },\tau _{1},t\right) }{\partial \mathbf{\pi }_{l}}-2\frac{%
\partial \mathbf{X}_{w}\left( \mathbf{\xi },\mathbf{\pi },\tau _{2},t\right) 
}{\partial \mathbf{\pi }_{l}}\right] \right\} _{\left\{ 
\begin{array}{c}
_{\mathbf{\xi }} \\ 
_{\mathbf{\pi }}%
\end{array}%
\right\} =\left\{ 
\begin{array}{l}
_{\mathbf{X},} \\ 
_{\mathbf{P}}%
\end{array}%
\right\} \left( \mathbf{X}\left( \mathbf{x},\mathbf{p},\tau _{1},0\right) ,%
\mathbf{P}\left( \mathbf{x},\mathbf{p},\tau _{1},0\right) +\hbar \mathbf{k}%
/2,t,\tau _{1}\right) }.  \label{49.2}
\end{eqnarray}%
When atoms move between the Raman pulses under the action of the homogeneous
gravitational field $\mathbf{g}$ of the Earth and the inhomogeneous
perturbation $\delta \mathbf{g}(\mathbf{x},t)$ caused by the test mass, the
only contribution to $\chi _{ikl}^{\prime }\left( \mathbf{\xi },t\right) $
[defined in Eq. (\ref{6b})] results from the presence of the test mass, 
\begin{subequations}
\begin{equation}
\chi _{ikl}^{\prime }\left( \mathbf{x},t\right) =M_{a}\chi _{ikl}\left( 
\mathbf{x},t\right) ,  \label{54b}
\end{equation}%
where 
\end{subequations}
\begin{equation}
\chi _{ijl}\left( \mathbf{x},t\right) =\partial _{\mathbf{x}_{l}}\gamma
_{ij}\left( \mathbf{x},t\right) .
\end{equation}

Since we calculate the AI phase to first order in $\delta \mathbf{g}$, it is
sufficient to calculate the atom trajectory in Eq. (\ref{49.2}) to zeroth
order in $\delta \mathbf{g}$, i.e. to set 
\begin{equation}
\left\{ 
\begin{array}{c}
\mathbf{X} \\ 
\mathbf{P}%
\end{array}%
\right\} \left( \mathbf{x},\mathbf{p},t,t^{\prime }\right) =\left\{ 
\begin{array}{c}
\mathbf{X}^{\left( 0\right) } \\ 
\mathbf{P}^{\left( 0\right) }%
\end{array}%
\right\} \left( \mathbf{x},\mathbf{p},t,t^{\prime }\right) ,  \label{55}
\end{equation}%
which results in 
\begin{equation}
\partial _{p_{i}}\mathbf{X}_{j}^{\left( 0\right) }\left( \mathbf{x},\mathbf{p%
},t,t^{\prime }\right) =\frac{\delta _{ij}}{M_{a}}\left( t-t^{\prime
}\right) ,  \label{56}
\end{equation}%
where $\delta _{ij}$ is a Kronecker delta. Moreover, since we are interested
in calculating $\phi _{Q}\left( \mathbf{x},\mathbf{p}\right) $ to second
order in the recoil momentum $\hbar \mathbf{k}$, we can neglect the
contribution of the recoil term in the braces of Eq. (\ref{49.2}). We then
apply the multiplication law (\ref{31}) and obtain 
\begin{equation}
\phi _{Q}\left( \mathbf{x},\mathbf{p}\right) =\dfrac{\hbar ^{2}}{24M_{a}^{2}}%
k_{i}k_{j}k_{l}\left\{ \int_{\tau _{1}}^{\tau _{2}}dt\chi _{ijl}\left[ 
\mathbf{X}^{\left( 0\right) }\left( \mathbf{x},\mathbf{p},t,0\right) ,t%
\right] \left( t-\tau _{1}\right) ^{3}+\int_{\tau _{2}}^{\tau _{3}}dt\chi
_{ijl}\left[ \mathbf{X}^{\left( 0\right) }\left( \mathbf{x},\mathbf{p}%
,t,0\right) ,t\right] \left( \tau _{3}-t\right) ^{3}\right\}  \label{58}
\end{equation}%
As before, we transform the integral to one from $0$ to $T$, 
\begin{equation}
\phi _{Q}\left( \mathbf{x},\mathbf{p}\right) =\dfrac{\hbar ^{2}}{24M_{a}^{2}}%
k_{i}k_{j}k_{l}\int_{0}^{T}d\theta \left\{ \theta ^{3}\chi _{ijl}\left[ 
\mathbf{X}^{\left( 0\right) }\left( \mathbf{x},\mathbf{p},\tau _{1}+\theta
,0\right) ,\tau _{1}+\theta \right] +\left( T-\theta \right) ^{3}\chi _{ijl}%
\left[ \mathbf{X}^{\left( 0\right) }\left( \mathbf{x},\mathbf{p},\tau
_{2}+\theta ,0\right) ,\tau _{2}+\theta \right] \right\} .  \label{59}
\end{equation}%
If the test mass moves without rotation, then%
\begin{equation}
\chi _{ikl}\left( \mathbf{x},t\right) =\chi _{ikl}\left[ \mathbf{x}-\mathbf{x%
}_{m}\left( t\right) \right]  \label{59.1}
\end{equation}%
and 
\begin{equation}
\phi _{Q}\left( \mathbf{x},\mathbf{p}\right) =\dfrac{\hbar ^{2}}{24M_{a}^{2}}%
k_{i}k_{j}k_{l}\int_{0}^{T}d\theta \left\{ 
\begin{array}{c}
\theta ^{3}\chi _{ijl}\left[ \mathbf{X}^{\left( 0\right) }\left( \mathbf{x},%
\mathbf{p},\tau _{1}+\theta ,0\right) -\mathbf{x}_{m}\left( \tau _{1}+\theta
\right) \right] \\ 
+\left( T-\theta \right) ^{3}\chi _{ijl}\left[ \mathbf{X}^{\left( 0\right)
}\left( \mathbf{x},\mathbf{p},\tau _{2}+\theta ,0\right) -\mathbf{x}%
_{m}\left( \tau _{2}+\theta \right) \right]%
\end{array}%
\right\} .  \label{61}
\end{equation}%
This is the exact expression for $\mathbf{\psi }_{q}$ that is used in Sec.
III.

We can obtain an approximate expression for $\phi _{Q}\left( \mathbf{x},%
\mathbf{p}\right) $ when the distance between the atoms and the test mass is
sufficiently large to keep only those terms that are linear in the field
curvature. If we choose%
\begin{equation}
\mathbf{x}_{QC}=\QDABOVE{1pt}{2}{T^{4}}\int_{0}^{T}d\theta \left\{ \theta
^{3}\left[ \mathbf{X}^{\left( 0\right) }\left( \mathbf{x},\mathbf{p},\tau
_{1}+\theta ,0\right) -\mathbf{x}_{m}\left( \tau _{1}+\theta \right) \right]
+\left( T-\theta \right) ^{3}\left[ \mathbf{X}^{\left( 0\right) }\left( 
\mathbf{x},\mathbf{p},\tau _{2}+\theta ,0\right) -\mathbf{x}_{m}\left( \tau
_{2}+\theta \right) \right] \right\}  \label{63}
\end{equation}%
and expand 
\begin{equation}
\chi _{ijl}\left( \mathbf{x}\right) \approx \chi _{ijl}\left( \mathbf{x}%
_{QC}\right) +\left( \partial _{\mathbf{x}_{C}}\right) _{m}\chi _{ijl}\left( 
\mathbf{x}_{QC}\right) \left( \mathbf{x}-\mathbf{x}_{QC}\right) _{m},
\label{62}
\end{equation}%
the contribution to Q-term from the second term in Eq. (\ref{62}) vanishes
and we find an approximate expression $\phi _{Qa}$ for the phase given by 
\begin{equation}
\phi _{Q}\left( \mathbf{x},\mathbf{p}\right) \approx \phi _{Qa}\left( 
\mathbf{x},\mathbf{p}\right) =\dfrac{\hbar ^{2}}{48M_{a}^{2}}%
k_{i}k_{j}k_{l}\chi _{ijl}\left( \mathbf{x}_{QC}\right) T^{4}.  \label{64}
\end{equation}

\section{\label{s3}Point source test mass}

For a point source test mass $M$ moving along the trajectory $\mathbf{x}%
_{m}\left( t\right) $, the gravitational field, gravity-gradient tensor, and
gravity curvature tensor are given by 
\begin{subequations}
\label{60}
\begin{eqnarray}
\delta \mathbf{g}\left( \mathbf{x},t\right) &=&\delta \mathbf{g}\left[ 
\mathbf{x}-\mathbf{x}_{m}\left( t\right) \right] \text{; \ \ \ }\delta 
\mathbf{g}\left( \mathbf{x}\right) =-GM\dfrac{\mathbf{x}}{x^{3}},
\label{60a} \\
\gamma _{jl}\left( \mathbf{x},t\right) &=&\gamma _{jl}\left[ \mathbf{x}-%
\mathbf{x}_{m}\left( t\right) \right] \text{; \ \ \ }\gamma _{jl}\left( 
\mathbf{x}\right) =-GM\left( \frac{\delta _{jl}}{x^{3}}-3\frac{x_{j}x_{l}}{%
x^{5}}\right) ,  \label{60b} \\
\chi _{ijl}\left( \mathbf{x},t\right) &=&\chi _{ijl}\left[ \mathbf{x}-%
\mathbf{x}_{m}\left( t\right) \right] ,  \label{60c} \\
\chi _{ijl}\left( \mathbf{x}\right) &=&GM\left[ \frac{3}{x^{5}}\left( \delta
_{jl}x_{i}+\delta _{il}x_{j}+\delta _{ij}x_{l}\right) -15\frac{%
x_{i}x_{j}x_{l}}{x^{7}}\right] ,  \label{60d}
\end{eqnarray}%
where $G=6.67428\times 10^{-11}$ m$^{3}/$kg$\cdot $s$^{2}$ is the Newtonian
gravitational constant.

We numerically calculated the classical part of the phase given by Eqs. (\ref%
{201d}, \ref{202.3}) (for $\delta \phi $) and the quantum corrections given
by Eqs. (\ref{201e}, \ref{207.1aaa}) (for $\phi _{q}$) and (\ref{61}) (for $%
\phi _{Q}$) and determined when these terms become measurable in the
presence of phase noise given by Eq. (\ref{2.2a}). Moreover, we checked the
validity of the approximate expressions given in Eqs. (\ref{202.7}, \ref%
{202.6}), (\ref{207.1bbb}), (\ref{211}, \ref{210}), and (\ref{64}, \ref{63}%
). The results vary with the test mass' weight, shape, trajectory, as well
as with the operating parameters of the atom interferometer.

The calculations are carried out for a test mass moving with constant
velocity $\mathbf{v}_{m}$, 
\end{subequations}
\begin{equation}
\mathbf{x}_{m}\left( t\right) =\mathbf{x}_{m0}+\mathbf{v}_{m}t,  \label{65}
\end{equation}%
where 
\begin{equation}
\mathbf{x}_{m0}=\left( x_{m0},y_{m0},z_{m0}\right) .  \label{67}
\end{equation}%
is the location of the test mass at time $t=0.$

We assume that, at $t=0$, the atoms are launched from the origin of the
North-East-Down frame in the vertical direction. That is, it is assumed that 
$\mathbf{g}$ is in the positive $z$ direction, and that the cloud position
is given by 
\begin{subequations}
\begin{eqnarray}
z(t) &=&v_{0}t+\dfrac{1}{2}gt^{2};  \label{c1a} \\
x(t) &=&y(t)=0,
\end{eqnarray}%
where $\mathbf{v}_{0}=v_{0}\mathbf{u}_{z}$ is the launch velocity, taken to
be along the $z-$axis. In this case one finds from Eqs. (\ref{101c}, \ref%
{202.6}, \ref{210}, \ref{63}) that $\ $%
\end{subequations}
\begin{subequations}
\label{66}
\begin{eqnarray}
\mathbf{x}_{C} &=&-\mathbf{x}_{m0}+\left[ \mathbf{v}_{0}-\mathbf{v}_{m}%
\right] \left( \tau _{1}+T\right) +\dfrac{1}{2}\mathbf{g}\left( \tau
_{1}^{2}+2\tau _{1}T+\dfrac{7}{6}T^{2}\right) ,  \label{66a} \\
\mathbf{x}_{qC} &=&-\mathbf{x}_{m0}+\left[ \mathbf{v}_{0}-\mathbf{v}_{m}%
\right] \left( \tau _{1}+\dfrac{7}{6}T\right) +\dfrac{1}{2}\mathbf{g}\left(
\tau _{1}^{2}+\dfrac{7}{3}\tau _{1}T+\dfrac{3}{2}T^{2}\right) ,  \label{66b}
\\
\mathbf{x}_{QC} &=&-\mathbf{x}_{m0}+\left[ \mathbf{v}_{0}-\mathbf{v}_{m}%
\right] \left( \tau _{1}+T\right) +\mathbf{g}\left( \frac{\tau _{1}^{2}}{2}%
+\tau _{1}T+\frac{8}{15}T^{2}\right) .  \label{66c}
\end{eqnarray}%
Recall that these values were chosen to insure that the first order terms
vanish in the expansions given in Eq. (\ref{202.4}), (\ref{208}), and (\ref%
{62}). As such, by expanding the expressions for $\psi $ about these points,
we obtained approximate expressions with corrections of order $\left\vert 
\mathbf{x}-\mathbf{x}_{C}\right\vert ^{2}$ times derivatives of the gravity
gradient tensor, derivatives of the gravity-curvature tensor, and second
derivatives of the gravity-curvature tensor for $\delta \phi $, $\phi _{q}$
and $\phi _{Q}$, respectively. As such, these choices improve the accuracy
of the approximations.

The launch velocity is chosen as 
\end{subequations}
\begin{equation}
\mathbf{v}_{0}=-\mathbf{g}\left( \tau _{1}+T\right) ,  \label{v1a}
\end{equation}%
corresponding to a symmetric fountain geometry in which the atomic cloud
reaches its highest point at time $\tau _{2}$, when the second pulse is
applied. Calculations have been performed for a stationary test mass having 
\begin{equation}
\mathbf{x}_{m}^{(1)}\left( t\right) =\mathbf{x}_{m0}=\left(
0,y_{m0},z_{m0}\right)
\end{equation}%
and a test mass moving with constant velocity $\mathbf{v}_{m}=\left( 5\text{
m/s,0,0}\right) $ along the $x$ axis, 
\begin{equation}
\mathbf{x}_{m}^{(2)}\left( t\right) =\mathbf{x}_{m0}+\mathbf{v}_{m}t=\left(
x_{m0},y_{m0},z_{m0}\right) +\left( 5\text{ m/s,0,0}\right) t.
\end{equation}%
The parameters characterizing the atom interferometer, the test mass, and
the Earth's field are summarized in Table \ref{t1}. 
\begin{table}[tbph]
\begin{tabular}{ll}
Earth's gravitational field & $\mathbf{g}=\left\{ 0,0,9.8\text{ m/s}^{\text{2%
}}\right\} $ \\ 
Multiple-$\hbar k$ beam splitter factor & $n_{k}=25$ \\ 
Effective wave vector & $\mathbf{k}=\left\{ 0,0,-k\right\} ,~k=4.0275\times
10^{8}$ m$^{-1}$ \\ 
Time between launch and first Raman pulse & $\tau _{1}=10$ ms \\ 
Time between Raman pulses & $T=200$ ms \\ 
Launch velocity & $\mathbf{v}_{0}=-\mathbf{g}\left( \tau _{1}+T\right) $ \\ 
Error of atom interferometer phase measurement & $\phi _{err}=10^{-3}$ rad
\\ 
test mass & $M=50$ kg \\ 
atomic mass & 87%
\end{tabular}%
\caption{Parameters of the atom interferometer and gravitational sources.}
\label{t1}
\end{table}
Since the cloud trajectory and the effective wave vectors are vertical, $%
y_{m0}$ can be considered as an "impact parameter" for the test mass
relative to the cloud trajectory along the $z-$axis

Equations (\ref{60}) can be used either for a point mass or a spherical mass
having constant density $\rho $ and radius%
\begin{equation}
y_{\min }=\left[ \frac{3M}{4\pi \rho }\right] ^{1/3}.  \label{q}
\end{equation}%
For the highest density in nature, $\rho =22600$ kg/m$^{3}$, corresponding
to osmium \cite{a35}, 
\begin{equation}
y_{\min }\approx 0.0808\text{ m.}  \label{qq}
\end{equation}%
For impact parameters $y_{m0}<y_{\min }$ Eqs. (\ref{60}) are valid only for
those values of $z_{m0}$ for which atom trajectory does not intersect the
spherical test mass. In the case of a stationary sphere, this requirement
translates into one in which the distance between the cloud and the sphere
is always greater than $y_{\min }$; that is, for any $t>0,$%
\begin{equation}
y_{m0}^{2}+\left[ z(t)-z_{m0}\right] ^{2}>y_{\min }^{2}.
\end{equation}%
With $z(t)$ given by Eq. (\ref{c1a}) and $\mathbf{v}_{0}$ given by Eq. (\ref%
{v1a}), we can show that this inequality is satisfied if%
\begin{equation}
-\sqrt{y_{\min }^{2}-y_{m0}^{2}}-\frac{1}{2}g\left( \tau _{1}+T\right)
^{2}>z_{m0},~\text{or}~z_{m0}>\sqrt{y_{\min }^{2}-y_{m0}^{2}}.  \label{q.1}
\end{equation}%
For the moving sphere the range of allowed initial positions $\left(
x_{m0},y_{m0},z_{m0}\right) $ for the center of the sphere is more difficult
to calculate. For this reason, in the case 2 we consider only impact
parameters larger than sphere's radius, $y_{m0}>y_{\min },$ for which,
evidently, any values of $\left( x_{m0},z_{m0}\right) $ are allowed.

\begin{figure}[!t]
\includegraphics[width=10.9cm]{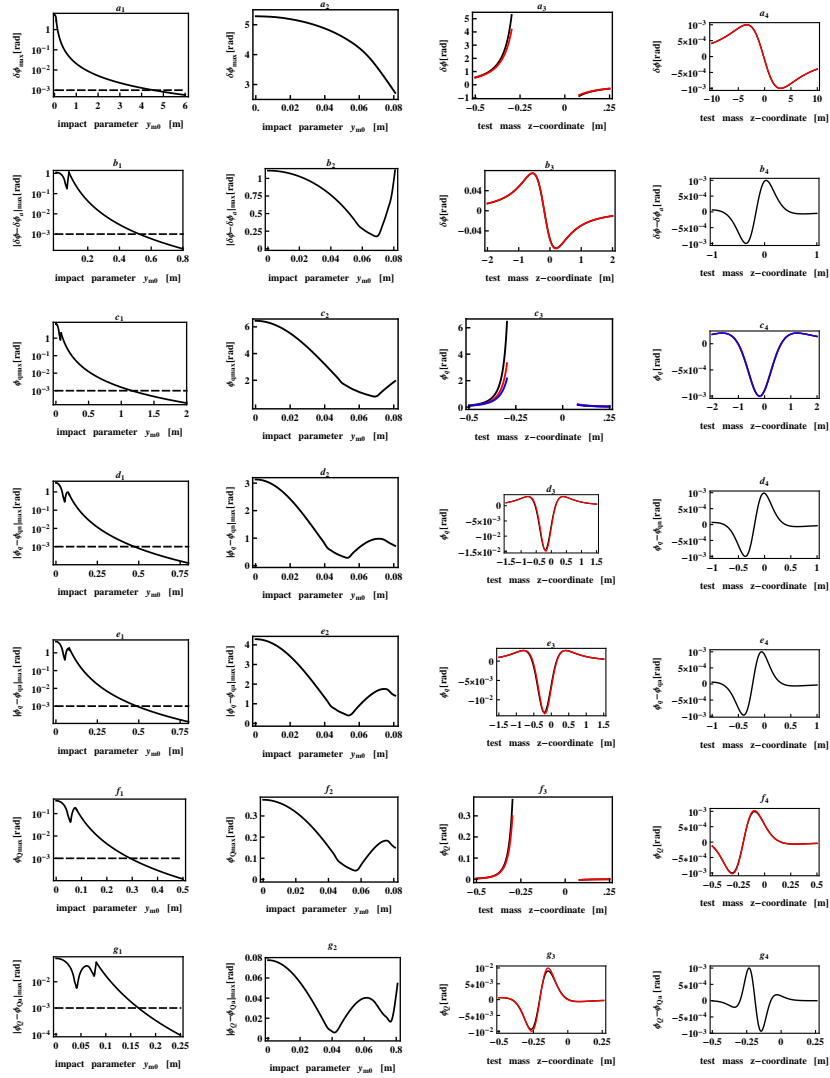}
\caption{Stationary source.\textbf{\ }Plot%
\textbf{\ }$a_{1}$, maximum of the phase magnitude given in Eqs. (\protect
\ref{201d}, \protect\ref{202.3}) versus the impact parameter $y_{m0}$; Plot $%
a_{2},$ the same as plot $a_{1}$ but for $y_{m0}<y_{\min }$; Plot $a_{3}$,
dependence of the exact (black curve) and approximate (red curve) phases as
a function of the initial $z-$coordinate ($z_{m0}$) for $\left\{
x_{m0},y_{m0}\right\} =\left\{ 0,0\right\} $ $,$ where the phase achieves
its maximum magnitude, Plot $a_{4}$, the same as plot $a_{3}$ but for $%
\left\{ x_{m0},y_{m0}\right\} =\left\{ 0,4.55\text{ m}\right\} ,$ where the
phase passes below the noise threshhold $\protect\phi _{err}.$ Plots\textbf{%
\ }$\left\{ b_{1},b_{2}\right\} $, the same as plot $\left\{
a_{1},a_{2}\right\} $ but for the maximum magnitude of the difference
between exact and approximate phases given in Eqs. (\protect\ref{201d}, 
\protect\ref{202.3}) and (\protect\ref{202.7}, \protect\ref{66a}); Plot $%
b_{3}$, the same as plot $a_{3}$, but for $\left\{ x_{m0},y_{m0}\right\}
=\left\{ 0,0.525\text{ m}\right\} ,$ where the magnitude of phase difference
shown on plot $b_{1}$ passes below $\protect\phi _{err};$ Plot $b_{4},$
difference between black and red curves in plot $b_{3}.$ Plots\textbf{\ }$%
\left\{ c_{1}-c_{4}\right\} $, the same as plots $\left\{
a_{1}-a_{4}\right\} $ but for the magnitude of the quantum correction given
in Eqs. (\protect\ref{201e}, \protect\ref{207.1aaa}); the values of $\left\{
x_{m0},y_{m0}\right\} $ are $\left\{ 0,0\right\} $ for $c_{3}$ and $\left\{
0,1.16\text{ m}\right\} $ for $c_{4}$, where the magnitude of the phase
difference shown on plot $c_{1}$ passes below $\protect\phi _{err}$; exact
quantum correction $\protect\phi _{q},$ approximations $\protect\phi _{qn}$
and $\protect\phi _{qa}$ are shown in black, red and blue, respectively.%
\textbf{\ }Plots\textbf{\ }$\left\{ d_{1}-d_{4}\right\} $, the same as plots 
$\left\{ b_{1}-b_{4}\right\} $ but for the maximum magnitude of the
difference between exact and approximate quantum corrections given in Eqs. (%
\protect\ref{201e}, \protect\ref{207.1aaa}) and (\protect\ref{206a}, \protect
\ref{207.1bbb}); $\left\{ x_{m0},y_{m0}\right\} =\left\{ 0,0.473\text{ m}%
\right\} $ in plot $d_{3}$ where the magnitude of the phase difference shown
on plot $d_{1}$ passes below $\protect\phi _{err}.$ Plots\textbf{\ }$\left\{
e_{1}-e_{4}\right\} $, the same as plots $\left\{ d_{1}-d_{4}\right\} $ but
for the maximum magnitude of the difference between exact and approximate
quantum corrections given in Eqs. (\protect\ref{201e}, \protect\ref{207.1aaa}%
) and (\protect\ref{211}, \protect\ref{66b}); $\left\{ x_{m0},y_{m0}\right\}
=\left\{ 0,0.488\text{ m}\right\} $ in plot $e_{3},$ where the magnitude of
the phase difference shown on plot $e_{1}$ passes below $\protect\phi _{err}$%
. Plots $\left\{ f_{1}-f_{4}\right\} $, the same as plots $\left\{
a_{1}-a_{4}\right\} $ but for the magnitude of the $Q-$term given in Eq. (%
\protect\ref{61}), values of $\left\{ x_{m0},y_{m0}\right\} $ are $\left\{
0,0\right\} $ for $f_{3}$ and $\left\{ x_{m0},y_{m0}\right\} =$ $\left\{
0,0.292\text{ m}\right\} $ for $f_{4},$ where the magnitude of the phase
shown on plot $f_{1}$ passes below $\protect\phi _{err}.$ Plots\textbf{\ }$%
\left\{ g_{1}-g_{4}\right\} $, the same as plots $\left\{
b_{1}-b_{4}\right\} $ but for the maximum magnitude of the difference
between exact and approximate $Q-$terms given in Eqs. (\protect\ref{61}) and
(\protect\ref{64}, \protect\ref{66c}); $\left\{ x_{m0},y_{m0}\right\}
=\left\{ 0,0.166\text{ m}\right\} $ in plot $g_{3}$, where the magnitude of
the difference shown on plot $g_{1}$ passes below $\protect\phi _{err}.$}
\label{f1}
\end{figure}

\begin{figure}[!t]
\includegraphics[width=10.9cm]{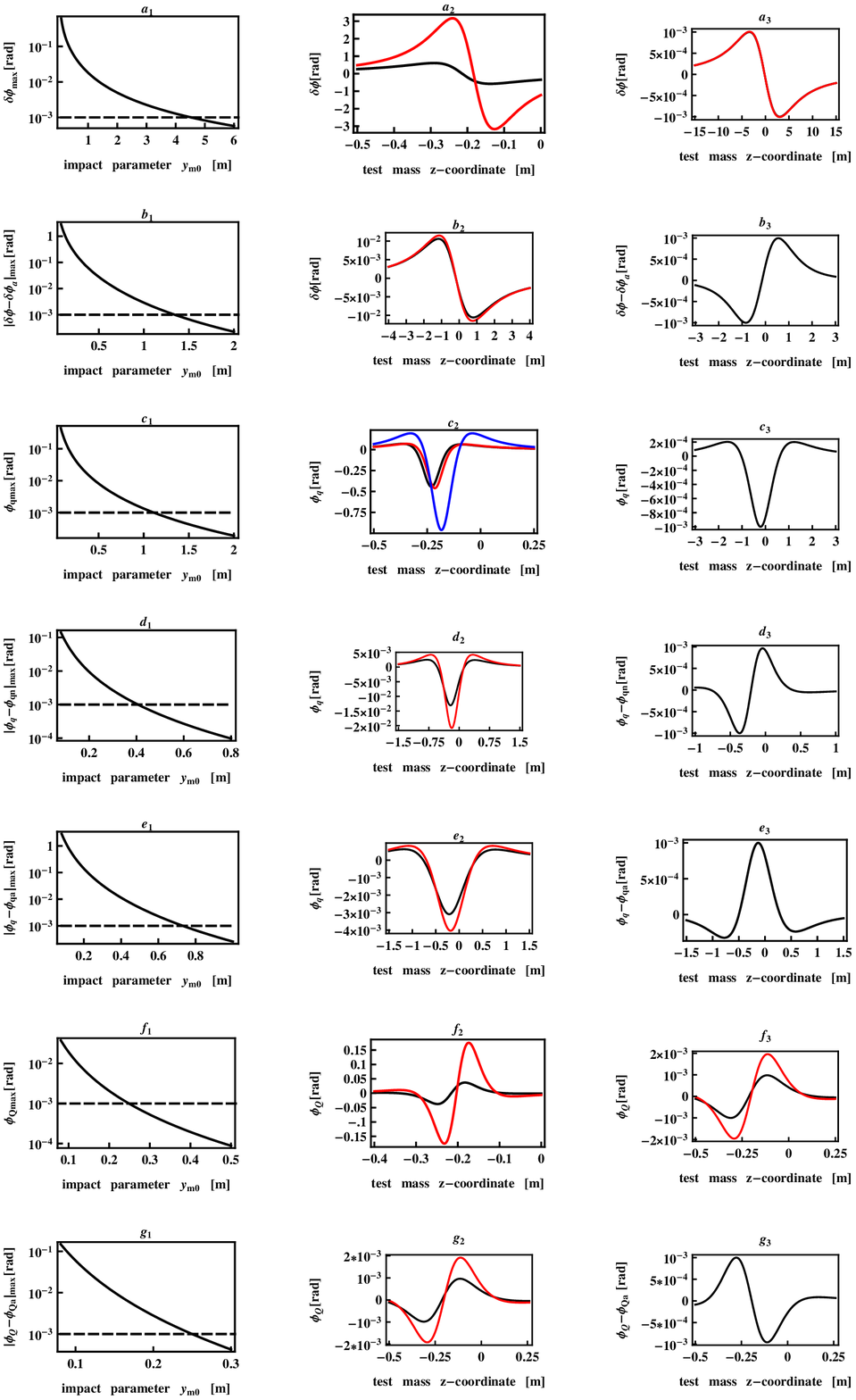}
\caption{Test mass moving with constant
velocity $5$ m/s. Columns 1, 2, and 3, mirror columns 1,3, and 4 of Fig. 
\protect\ref{f1}$.$ Values of $\left\{ x_{m0},y_{m0}\right\} $ are $\left\{
-1.05\text{ m},y_{\min }\right\} $ for plots $a_{2},$ $f_{2},$ $\left\{ -1.05%
\text{ m},4.53\text{ m}\right\} $ for plot $a_{3}$, $\left\{
-1.05,1.34\right\} $ for plots $b_{2},\left\{ -1.13\text{ m},y_{\min
}\right\} $ for plot $c_{2},$ $\left\{ -1.22\text{ m},1.11\text{ m}\right\} $
for plot $c_{3},\left\{ -1.33\text{ m},0.407\text{ m}\right\} $ for plot $%
d_{2},$ $\left\{ -1.21\text{ m},0.732\text{ m}\right\} $ for plot $e_{2},$ $%
\left\{ -1.05\text{ m},0.249\text{ m}\right\} $ for plot $f_{3},$ and $%
\left\{ -1.05\text{ m},0.250\text{ m}\right\} $ for plot $g_{2}.$}
\label{f2}
\end{figure}

For each impact parameter $y_{m0}$ and for the parameters given in Table \ref%
{t1}$,$ we explore various test mass positions and trajectories. The results
of the calculations are illustrated graphically in Fig. \ref{f1} for a
stationary mass and in Fig. \ref{f2} for a mass moving at constant velocity.
Although the various plots may be difficult to read at standard
magnification, they can be read easily using the zoom feature when read
online in PDF format.

In the first two columns of Fig. \ref{f1}, we plot

\begin{enumerate}
\item maximum of the magnitude of the phase $\left\vert \delta \phi
\right\vert _{\max }$ obtained from Eqs. (\ref{201d}, \ref{202.3}), Plots a$%
_{1},$a$_{2};$

\item maximum of the magnitude of the phase difference $\left\vert \delta
\phi -\delta \phi _{a}\right\vert _{\max }$ obtained from Eqs. (\ref{201d}, %
\ref{202.3}; \ref{202.7}, \ref{66a}), Plots b$_{1},$b$_{2};$

\item the maximum of the magnitude of the quantum correction $\left\vert
\phi _{q}\right\vert _{\max }$ obtained from Eqs. (\ref{201e}, \ref{207.1aaa}%
), Plots c$_{1},$c$_{2};$

\item maximum of the magnitude of the phase difference $\left\vert \phi
_{q}-\phi _{qn}\right\vert _{\max }$ obtained from Eqs. (\ref{201e}, \ref%
{207.1aaa}; \ref{206a}, \ref{207.1bbb}), Plots d$_{1},$d$_{2};$

\item maximum of the magnitude of the phase difference $\left\vert \phi
_{q}-\phi _{qa}\right\vert _{\max }$ obtained from Eqs. (\ref{201e}, \ref%
{207.1aaa}; \ref{211}, \ref{66b}), Plots e$_{1},$e$_{2}$;

\item maximum of the magnitude of the quantum correction $\left\vert \phi
_{Q}\right\vert _{\max }$ obtained from Eq. (\ref{61}), Plots f$_{1},$f$%
_{2}; $

\item maximum of the magnitude of the phase difference $\left\vert \phi
_{Q}-\phi _{Qa}\right\vert _{\max }$ obtained from Eqs. (\ref{61}; \ref{64}, %
\ref{66c}), Plots g$_{1},$g$_{2}$.
\end{enumerate}

In effect, column 2 is a blow-up of column 1 for values of $y_{m0}<y_{\min
}. $ Values of $y_{m0}<y_{\min }$ are allowed provided inequalities (\ref%
{q.1}) holds. For the stationary test mass, the maximum values of the
various phases and phase differences occur if the mass is positioned as
close as possible to the top of the cloud trajectory, without touching it.
For the parameters given in Table \ref{t1} and the trajectory determined by
Eqs. (\ref{c1a}) and (\ref{v1a}), the top of the cloud trajectory occurs for 
$z_{\min }=-0.22$ m. As a consequence the maximum phases occur for $y_{m0}=0$%
, $z_{m0}=z_{\min }-y_{\min }=-0.30$ m. That the maximum phases occur for $%
y_{m0}=0$ is evident in column 2. The plots in column 1 of Fig. \ref{f2}
mirror those of column 1 of Fig. \ref{f1}, except that Fig. \ref{f2} is
drawn for a test mass moving with constant velocity in Fig. \ref{f2}. The
maximum phases in this case occur for $y_{m0}\approx y_{\min }$.

Phases $\delta \phi $, $\phi _{q}$, and $\phi _{Q}$ that lie above the
dashed lines in these plots are measurable since they exceed the noise
level. On the other hand, phase \textit{differences} between the exact and
approximate results must lie \textit{below} the dashed lines for the
approximations to be good. For example, in plot a$_{1}$ we see that the
signal exceeds the noise only if $y_{m0}<4.55$ m and, in plot b$_{1}$, we
see that we see that the difference between the exact and approximate
expressions is below the noise level only if $y_{m0}>0.525$ m$.$ By
examining the plots in column 1 of Figs. \ref{f1} and. \ref{f2}, we able to
determine the regions in which the interferometric signal rises above the
noise and also to determine the range of validity of the various
approximation expressions that we derived. The results are summarized below
for the regions of $y_{m0}$ listed in Table \ref{t2} that were obtained from
column 1 of Figs. \ref{f1} and. \ref{f2}.

\begin{table}[tbph]
\begin{tabular}{|c|c|c|}
\hline
Region & stationary test mass & test mass moving with constant velocity \\ 
\hline
1 & $y_{m0}<0.166$ m & $y_{m0}<0.250$ m \\ \hline
2 & $y_{m0}>0.292$ m & $y_{m0}>0.249$ m \\ \hline
3 & $y_{m0}<0.473$ m & $y_{m0}<0.407$ m \\ \hline
4 & $y_{m0}>0.488$ m & $y_{m0}>0.732$ m \\ \hline
5 & $y_{m0}<0.525$ m & $y_{m0}<1.34$ m \\ \hline
6 & $y_{m0}>1.16$ m & $y_{m0}>1.11$ m \\ \hline
7 & $y_{m0}>4.55$ m & $y_{m0}>4.53$ m \\ \hline
\end{tabular}%
\caption{Locations of regions of validity of the exact and approximate
expressions for the stationary and moving test mss. The regions refer to
regions 1-7 given in the text.}
\label{t2}
\end{table}

\begin{enumerate}
\item[Region 1.] One should use the exact expression, Eq. (\ref{61}), for $%
\phi _{Q}$ in this region$;$ only outside this region are the approximate
expressions given by Eqs. (\ref{64}, \ref{66c}) valid (see plots $g_{1}$ in
the figures);

\item[Region 2.] The phase $\phi _{Q}$ is negligible in this region (see
plots $f_{1}$ in the figures);

\item[Region 3.] One should use the exact expressions, Eqs. (\ref{201e}, \ref%
{207.1aaa}), for the quantum correction $\phi _{q}$ in this region$;$ only
outside this region does the approximate expression given by Eqs. (\ref{206a}%
, \ref{207.1bbb}) become valid (see plots $d_{1}$ in the figures);

\item[Region 4.] One can use the approximate expressions for $\phi _{qa}$
given by Eqs. (\ref{211}, \ref{66b}) in this region (see plots $e_{1}$ in
the figures);

\item[Region 5.] One should use the exact expressions, Eqs. (\ref{201d}, \ref%
{202.3}), for the classical part of the phase $\delta \phi ;$ only outside
this region does the approximate expression given by Eqs. (\ref{202.7}, \ref%
{66a}) become valid (see plots $b_{1}$ in the figures);

\item[Region 6.] The phase $\phi _{q}$ is negligible in this region (see
plots $c_{1}$ in the figures);

\item[Region 7.] The phase $\delta \phi $ produced by the test mass falls
below the phase noise $\phi _{err}$, so the effect of the test mass cannot
be measured in this region (see plots $a_{1}$ in the figures).
\end{enumerate}

Column 3 and 4 of Figs. \ref{f1} and. \ref{f2} give the dependence of the
various phases and phase corrections as a function of the initial $z-$%
coordinate, $z_{m0}$, for fixed $y_{m0}$. The value of $y_{m0}$ is chosen
either at a value that gives the maximum phase or at a value where the phase
crosses the noise threshold. with the value of $y_{m0}$ chosen that gives
rise to the largest phase or phase difference. For example, in plots a$_{3}$%
, c$_{3},$ and f$_{3},$ drawn for $y_{m0}=0$, we see that the maximum phases 
$\delta \phi ,$ $\phi _{q}$, and $\phi _{Q}$ occur $z_{m0}=-0.30$, when the
test mass is just above the zenith of the trajectory. Moreover $y_{m0}=0$,
it follows from inequalities (\ref{q.1}) that $z_{m0}<-0.30$ m and $%
z_{m0}>.08$ m. All these features are seen in plots a$_{3}$, c$_{3},$ and f$%
_{3}$. Similar considerations apply for all the other plots in columns 3 and
4 of Figs. \ref{f1} and. \ref{f2}.

These numerical calculations show that the approximate expressions that were
obtained based on assumptions about the approximate homogeneity of the field
in Ref. \cite{a34} for the classical contributions to the phase or on the
approximate homogeneity of the field gradient and curvature for the quantum
corrections to the phase become valid only in regions where the phases are
smaller by 1-2 orders of magnitude the maximum values for the phases that
occur in regions where the field inhomogeneity plays an important role.

By using an effective wave vector that is 25 times larger than the wave
vector of a two-field Raman pulse, we reach a limit where the quantum
correction $\phi _{q}$ is comparable with the classical part of the phase,
while the $Q-$term is still small. Further increase of the effective wave
vector or time interval $T$ between the pulses could bring us to situation
when quantum correction dominate over the classical part of the phase.

\appendix

\section{~~}

In this appendix, we show how we arrive at Eqs. (\ref{201}) from Eq. (\ref%
{49b}) and Eqs. (\ref{101}).

To calculate the phase $\phi _{r}\left( \mathbf{x},\mathbf{p}\right) $, we
need an approximate expression for the propagator $\mathbf{X}\left( \mathbf{X%
}\left( \mathbf{x},\mathbf{p},\tau _{1},0\right) ,\mathbf{P}\left( \mathbf{x}%
,\mathbf{p},\tau _{1},0\right) +\dfrac{\hbar \mathbf{k}}{2},\tau _{s},\tau
_{1}\right) .$ Using Eqs. (\ref{101}, \ref{102}) we find 
\begin{gather}
\mathbf{X}\left( \mathbf{X}\left( \mathbf{x},\mathbf{p},\tau _{1},0\right) ,%
\mathbf{P}\left( \mathbf{x},\mathbf{p},\tau _{1},0\right) +\dfrac{\hbar 
\mathbf{k}}{2},\tau _{s},\tau _{1}\right) \approx \mathbf{X}^{\left(
0\right) }\left( 
\begin{array}{c}
\mathbf{X}^{\left( 0\right) }\left( \mathbf{x},\mathbf{p},\tau _{1},0\right)
+\delta \mathbf{X}\left( \mathbf{x},\mathbf{p},\tau _{1},0\right) , \\ 
\mathbf{P}^{\left( 0\right) }\left( \mathbf{x},\mathbf{p},\tau _{1},0\right)
+\delta \mathbf{P}\left( \mathbf{x},\mathbf{p},\tau _{1},0\right) +\dfrac{%
\hbar \mathbf{k}}{2},\tau _{s},\tau _{1}%
\end{array}%
\right)  \notag \\
+\delta \mathbf{X}\left( \mathbf{X}^{\left( 0\right) }\left( \mathbf{x},%
\mathbf{p},\tau _{1},0\right) ,\mathbf{P}^{\left( 0\right) }\left( \mathbf{x}%
,\mathbf{p},\tau _{1},0\right) +\dfrac{\hbar \mathbf{k}}{2},\tau _{s},\tau
_{1}\right)  \notag \\
=\mathbf{X}^{\left( 0\right) }\left( \mathbf{x},\mathbf{p},\tau
_{1},0\right) +\delta \mathbf{X}\left( \mathbf{x},\mathbf{p},\tau
_{1},0\right) +  \notag \\
\dfrac{1}{M_{a}}\left[ \mathbf{P}^{\left( 0\right) }\left( \mathbf{x},%
\mathbf{p},\tau _{1},0\right) +M_{a}\int_{0}^{\tau _{1}}dt\delta \mathbf{g}%
\left[ \mathbf{X}^{\left( 0\right) }\left( \mathbf{x},\mathbf{p},t,0\right)
,t\right] +\dfrac{\hbar \mathbf{k}}{2}\right] \left( \tau _{s}-\tau
_{1}\right) +\dfrac{1}{2}\mathbf{g}\left( \tau _{s}-\tau _{1}\right) ^{2} 
\notag \\
+\int_{\tau _{1}}^{\tau _{s}}dt\left( \tau _{s}-t\right) \delta \mathbf{g}%
\left[ \mathbf{X}^{\left( 0\right) }\left( \mathbf{X}^{\left( 0\right)
}\left( \mathbf{x},\mathbf{p},\tau _{1},0\right) ,\mathbf{P}^{\left(
0\right) }\left( \mathbf{x},\mathbf{p},\tau _{1},0\right) +\dfrac{\hbar 
\mathbf{k}}{2},t,\tau _{1}\right) ,t\right] ,  \label{103}
\end{gather}%
From Eqs. (\ref{102}, \ref{101c}), it follows that%
\begin{eqnarray}
\mathbf{X}^{\left( 0\right) }\left( \mathbf{x},\mathbf{p},\tau _{s},0\right)
&=&\mathbf{X}^{\left( 0\right) }\left( \mathbf{X}^{\left( 0\right) }\left( 
\mathbf{x},\mathbf{p},\tau _{1},0\right) ,\mathbf{P}^{\left( 0\right)
}\left( \mathbf{x},\mathbf{p},\tau _{1},0\right) ,\tau _{s},\tau _{1}\right)
\notag \\
&=&\mathbf{X}^{\left( 0\right) }\left( \mathbf{x},\mathbf{p},\tau
_{1},0\right) +\dfrac{1}{M_{a}}\mathbf{P}^{\left( 0\right) }\left( \mathbf{x}%
,\mathbf{p},\tau _{1},0\right) \left( \tau _{s}-\tau _{1}\right) +\dfrac{1}{2%
}\mathbf{g}\left( \tau _{s}-\tau _{1}\right) ^{2},  \label{104}
\end{eqnarray}%
allowing us to rewrite%
\begin{gather}
\mathbf{X}\left( \mathbf{X}\left( \mathbf{x},\mathbf{p},\tau _{1},0\right) ,%
\mathbf{P}\left( \mathbf{x},\mathbf{p},\tau _{1},0\right) +\dfrac{\hbar 
\mathbf{k}}{2},\tau _{s},\tau _{1}\right) \approx \mathbf{X}^{\left(
0\right) }\left( \mathbf{x},\mathbf{p},\tau _{s},0\right) +\dfrac{\hbar 
\mathbf{k}}{2M_{a}}\left( \tau _{s}-\tau _{1}\right)  \notag \\
+\left( \tau _{s}-\tau _{1}\right) \int_{0}^{\tau _{1}}dt\delta \mathbf{g}%
\left[ \mathbf{X}^{\left( 0\right) }\left( \mathbf{x},\mathbf{p},t,0\right)
,t\right] +\int_{0}^{\tau _{1}}dt\left( \tau _{1}-t\right) \delta \mathbf{g}%
\left[ \mathbf{X}^{\left( 0\right) }\left( \mathbf{x},\mathbf{p},t,0\right)
,t\right]  \notag \\
+\int_{\tau _{1}}^{\tau _{s}}dt\left( \tau _{s}-t\right) \delta \mathbf{g}%
\left[ \mathbf{X}^{\left( 0\right) }\left( \mathbf{X}^{\left( 0\right)
}\left( \mathbf{x},\mathbf{p},\tau _{1},0\right) ,\mathbf{P}^{\left(
0\right) }\left( \mathbf{x},\mathbf{p},\tau _{1},0\right) +\dfrac{\hbar 
\mathbf{k}}{2},t,\tau _{1}\right) ,t\right] .  \label{105}
\end{gather}

Using the relationship%
\begin{gather}
\mathbf{X}^{\left( 0\right) }\left( \mathbf{X}^{\left( 0\right) }\left( 
\mathbf{x},\mathbf{p},\tau _{1},0\right) ,\mathbf{P}^{\left( 0\right)
}\left( \mathbf{x},\mathbf{p},\tau _{1},0\right) +\dfrac{\hbar \mathbf{k}}{2}%
,t,\tau _{1}\right) =\dfrac{\hbar \mathbf{k}}{2M_{a}}\left( t-\tau
_{1}\right)  \notag \\
+\mathbf{X}^{\left( 0\right) }\left( \mathbf{X}^{\left( 0\right) }\left( 
\mathbf{x},\mathbf{p},\tau _{1},0\right) ,\mathbf{P}^{\left( 0\right)
}\left( \mathbf{x},\mathbf{p},\tau _{1},0\right) ,t,\tau _{1}\right)  \notag
\\
=\mathbf{X}^{\left( 0\right) }\left( \mathbf{x},\mathbf{p},t,0\right) +%
\dfrac{\hbar \mathbf{k}}{2M_{a}}\left( t-\tau _{1}\right) ,  \label{106}
\end{gather}%
we can write%
\begin{gather}
\mathbf{X}\left( \mathbf{X}\left( \mathbf{x},\mathbf{p},\tau _{1},0\right) ,%
\mathbf{P}\left( \mathbf{x},\mathbf{p},\tau _{1},0\right) +\dfrac{\hbar 
\mathbf{k}}{2},\tau _{s},\tau _{1}\right) \approx \mathbf{X}^{\left(
0\right) }\left( \mathbf{x},\mathbf{p},\tau _{s},0\right) +\dfrac{\hbar 
\mathbf{k}}{2M_{a}}\left( \tau _{s}-\tau _{1}\right)  \notag \\
+\left( \tau _{s}-\tau _{1}\right) \int_{0}^{\tau _{1}}dt\delta \mathbf{g}%
\left[ \mathbf{X}^{\left( 0\right) }\left( \mathbf{x},\mathbf{p},t,0\right)
,t\right] +\int_{0}^{\tau _{1}}dt\left( \tau _{1}-t\right) \delta \mathbf{g}%
\left[ \mathbf{X}^{\left( 0\right) }\left( \mathbf{x},\mathbf{p},t,0\right)
,t\right]  \notag \\
+\int_{\tau _{1}}^{\tau _{s}}dt\left( \tau _{s}-t\right) \delta \mathbf{g}%
\left[ \mathbf{X}^{\left( 0\right) }\left( \mathbf{x},\mathbf{p},t,0\right)
,t\right]  \notag \\
+\int_{\tau _{1}}^{\tau _{s}}dt\left( \tau _{s}-t\right) \left\{ \delta 
\mathbf{g}\left[ \mathbf{X}^{\left( 0\right) }\left( \mathbf{x},\mathbf{p}%
,t,0\right) +\dfrac{\hbar \mathbf{k}}{2M_{a}}\left( t-\tau _{1}\right) ,t%
\right] -\delta \mathbf{g}\left[ \mathbf{X}^{\left( 0\right) }\left( \mathbf{%
x},\mathbf{p},t,0\right) ,t\right] \right\} .  \label{107}
\end{gather}

The propagator (\ref{101e}) can be expressed as%
\begin{gather}
\delta \mathbf{X}\left( \mathbf{x},\mathbf{p},\tau _{s},0\right) \equiv
\int_{0}^{\tau _{s}}dt\left( \tau _{s}-t\right) \delta \mathbf{g}\left[ 
\mathbf{X}^{\left( 0\right) }\left( \mathbf{x},\mathbf{p},t,0\right) ,t%
\right]  \notag \\
=\int_{\tau _{1}}^{\tau _{s}}dt\left( \tau _{s}-t\right) \delta \mathbf{g}%
\left[ \mathbf{X}^{\left( 0\right) }\left( \mathbf{x},\mathbf{p},t,0\right)
,t\right] +\int_{0}^{\tau _{1}}dt\left( \tau _{s}-t\right) \delta \mathbf{g}%
\left[ \mathbf{X}^{\left( 0\right) }\left( \mathbf{x},\mathbf{p},t,0\right)
,t\right]  \notag \\
=\int_{\tau _{1}}^{\tau _{s}}dt\left( \tau _{s}-t\right) \delta \mathbf{g}%
\left[ \mathbf{X}^{\left( 0\right) }\left( \mathbf{x},\mathbf{p},t,0\right)
,t\right] +\int_{0}^{\tau _{1}}dt\left( \tau _{1}-t\right) \delta \mathbf{g}%
\left[ \mathbf{X}^{\left( 0\right) }\left( \mathbf{x},\mathbf{p},t,0\right)
,t\right]  \notag \\
+\left( \tau _{s}-\tau _{1}\right) \int_{0}^{\tau _{1}}dt\delta \mathbf{g}%
\left[ \mathbf{X}^{\left( 0\right) }\left( \mathbf{x},\mathbf{p},t,0\right)
,t\right] ,  \label{108}
\end{gather}%
which coincides with the sum of the 3rd, 4th and 5th terms on the
right-hand-side (rhs) of Eq. (\ref{107}) and reduces to 
\begin{gather}
\mathbf{X}\left( \mathbf{X}\left( \mathbf{x},\mathbf{p},\tau _{1},0\right) ,%
\mathbf{P}\left( \mathbf{x},\mathbf{p},\tau _{1},0\right) +\dfrac{\hbar 
\mathbf{k}}{2},\tau _{s},\tau _{1}\right) \approx \mathbf{X}^{\left(
0\right) }\left( \mathbf{x},\mathbf{p},\tau _{s},0\right) +\dfrac{\hbar 
\mathbf{k}}{2M_{a}}\left( \tau _{s}-\tau _{1}\right) +\delta \mathbf{X}%
\left( \mathbf{x},\mathbf{p},\tau _{s},0\right)  \notag \\
+\int_{\tau _{1}}^{\tau _{s}}dt\left( \tau _{s}-t\right) \left\{ \delta 
\mathbf{g}\left[ \mathbf{X}^{\left( 0\right) }\left( \mathbf{x},\mathbf{p}%
,t,0\right) +\dfrac{\hbar \mathbf{k}}{2M_{a}}\left( t-\tau _{1}\right) ,t%
\right] -\delta \mathbf{g}\left[ \mathbf{X}^{\left( 0\right) }\left( \mathbf{%
x},\mathbf{p},t,0\right) ,t\right] \right\} .  \label{200}
\end{gather}%
The first term on the right hand side of this equation is responsible for
the phase produced by the Earth's gravitational field. The second term
corresponds to the recoil correction to the first term, but this
contribution vanishes when Eq. (\ref{200}) is substituted into Eq. (\ref{49b}%
); there is no quantum correction in an homogeneous field. The third term is
responsible for the classical part of the phase produced by the test mass
while the fourth term is the recoil correction to the third term.

Substituting this result in the brackets of Eq. (\ref{49b}) for the 1st $%
\left( s=3\right) $ and 2nd $\left( s=2\right) $ terms, we arrive at Eqs. (%
\ref{201}).

\textbf{Acknowledgments}

The authors would like to thank Drs. M. Kasevich, M. Shverdin, B. Young, M.
Matthews, T. Loftus, V. Sonnad and A. Zorn for helpful discussions. This
work was performed under the auspices of the Department of Homeland
Security, Defense Nuclear Detection Office, Exploratory Research,
CFP11-100-RTA-06-FP007, Contract HSHQDC-11-X-00550 by Lawrence Livermore
National Laboratory, and the Defense Threat Reduction Agency by AOSense,
Inc., under Contract HDTRA1-13-C-0047. Lawrence Livermore National
Laboratory is operated by Lawrence Livermore National Security, LLC, for the
U.S. Department of Energy, National Nuclear Security Administration under
Contract DE-AC52-07NA27344

\textbf{Author Contributions}

S. Libby and B. Dubetsky considered the recoil term and carried out
numerical calculations of the classical and recoil contributions to the
phase. P. Berman and B. Dubetsky derived and interpreted the various
expressions for the phases and wrote the manuscript.

\textbf{Conflicts of Interest}

The authors declare that there are no conflicts of interest.


\begin{thebibliography}{99}
\bibitem{a1} Dubetsky, B.; Kazantsev, A. P.; Chebotayev, V. P.; Yakovlev, V.
P. Interference of atoms and formation of atomic spatial arrays in light
fields.\emph{\ Pis'ma Zh. Eksp. Teor. Fiz.} \textbf{1984}, \emph{39},
531-533 [\emph{JETP Lett.}\textbf{198}4, \emph{39}, 649-651].

\bibitem{a2} Weiss, D. S.; Young; B. C.; Chu, S. Precision Measurement of
the Photon Recoil of an Atom Using Atomic Interferometry. \emph{Phys. Rev.
Lett. }\textbf{1993}, \emph{70}, 2706-2709.

\bibitem{a3} Estey, B.; Yu, Ch.; M\"{u}ller, H.; Kuan, P.-C.; Lan, S.-Y.
High-resolution atom interferometers with suppressed diffraction phases.
http://arxiv.org/abs/1410.8486.

\bibitem{a4} Fixler, J. B.; Foster, G. T.; McGuirk, J. M.; Kasevich, M. A.
Atom Interferometer Measurement of the Newtonian Constant of Gravity. \emph{%
Science} \textbf{2007}, \emph{315}, 74-77.

\bibitem{a5} Rosi, G.; Sorrentino, F.; Cacciapuoti, L.; Prevedelli, M.;
Tino, G. M. Precision measurement of the Newtonian gravitational constant
using cold atoms. \emph{Nature} \textbf{2014}, \emph{510}, 518-521.

\bibitem{a6} Kasevich, M.; Chu, S. Atomic Interferometry Using Stimulated
Raman Transitions. \emph{Phys. Rev. Lett.} \textbf{1991}, \emph{67}, 181-184.

\bibitem{a6.1} Cahn, S. B.; Kumarakrishnan, A.; Shim, U.; Sleator, T.;
Berman, P. R.; Dubetsky, B. Time-Domain de Broglie Wave Interferometry. 
\emph{Phys. Rev. Lett.} \textbf{1997}, \emph{79}, 784-787.

\bibitem{a7} Peters, A.; Chung, K. Y.; Chu, S. Measurement of gravitational
acceleration by dropping atoms. \emph{Nature} \textbf{1999}, \emph{400},
849-852.

\bibitem{a7.1} Mok, C.; Barrett, B.; Carew, A.; Berthiaume, R.; Beattie, S.;
Kumarakrishnan A. Demonstration of improved sensitivity of echo
interferometers to gravitational acceleration.\emph{\ Phys. Rev. A }\textbf{%
2013}, \emph{88}, 023614.

\bibitem{a8} Snadden, M. J.; McGuirk, J. M.; Bouyer, P.; Haritos, K. G.;
Kasevich, M. A. Measurement of the Earth's Gravity Gradient with an Atom
Interferometer-Based Gravity Gradiometer. \emph{Phys. Rev. Lett.} \textbf{%
1998}, \emph{81}, 971-974.

\bibitem{a9} McGuirk, J. M.; Foster, G. T.; Fixler, J. B.; Snadden, M. J.;
Kasevich, M. A. Sensitive absolute-gravity gradiometry using atom
interferometry.\emph{\ Phys. Rev. A }\textbf{2002}, \emph{65}, 033608.

\bibitem{a9.1} Rosi, G.; Cacciapuoti, L.; Sorrentino, F.; Menchetti, M.;
Prevedelli, M.; Tino, G. M. Measurement of the Gravity-Field Curvature by
Atom Interferometry. \emph{Phys. Rev. Lett.} \textbf{2015}, \emph{114},
013001.

\bibitem{a10} Riehle, F.; Kisters, Th.; Witte, A.; Helmcke, J.; Borde, Ch.
J. Optical Ramsey Spectroscopy in a Rotating Frame: Sagnac Effect in a
Matter-Wave Interferometer. \emph{Phys. Rev. Lett.} \textbf{1991}, \emph{67}%
, 177-180.

\bibitem{a11} Lenef, A.; Hammond, T. D.; Smith, E. T.; Chapman, M. S.;
Rubenstein, R. A.; Pritchard, D. E. Rotation Sensing with an Atom
Interferometer. \emph{Phys. Rev. Lett. }\textbf{1997}, \emph{78}, 760-763.

\bibitem{a12} Gustavson, T. L.; Bouyer, P.; Kasevich, M. A. Precision
Rotation Measurements with an Atom Interferometer Gyroscope. \emph{Phys.
Rev. Lett. }\textbf{1997}, \emph{78}, 2046-2049.

\bibitem{a13} Canuel, B.; Leduc, F.; Holleville, D.; Gauguet, A.; Fils, J.;
Virdis, A.; Clairon, A.; Dimarcq, N.; Borde, Ch. J.; Landragin, A. Six-Axis
Inertial Sensor Using Cold-Atom Interferometry.\emph{\ Phys. Rev. Lett.} 
\textbf{2006}, \emph{97}, 010402.

\bibitem{a14} Dubetsky, B.; Kasevich, M. A. Atom interferometer as a
selective sensor of rotation or gravity. \emph{Phys. Rev. A }\textbf{2006}, 
\emph{74}, 023615.

\bibitem{a15} Barrett, B.; Geiger, R.; Dutta, I.; Meunier, M.; Canuel, B.;
Gauguet, A.; Bouyer, P.; Landragin, A. The Sagnac effect: 20 years of
development in matter-wave interferometry. \emph{C. R. Physique }\textbf{2014%
}, \emph{15}\textbf{,} 875-883.

\bibitem{a15.1} Wu, B.; Wang, Z Y ; Cheng, B.; Wang, Q. Y.; Xu, A. P.; Lin,
Q.\ Accurate measurement of the quadratic Zeeman coefficient of 87Rb clock
transition based on the Ramsey atom interferometer. \emph{J. Phys. B: At.
Mol. Opt. Phys.} \textbf{2014,} \emph{47,} 015001 (5pp)

\bibitem{a16} Biedermann, G. W.; Wu, X.; Deslauriers, L.; Roy, S.;
Mahadeswaraswamy, C.; Kasevich, M. A. Testing Gravity with Cold-Atom
Interferometers. http://arxiv.org/abs/1412.3210.

\bibitem{a17} Hamilton, P.; Jaffe, M.; Haslinger, P.; Simmons, Q.; M\"{u}%
ller, H. Atom-Interferometry Constraints on Dark Energy.
http://arxiv.org/abs/1502.03888

\bibitem{a18} Borde, C. J.; Howard, J-C.; Karasiewicz, A. Relativistic phase
shifts for Dirac particles interacting with weak gravitational fields in
matter-wave interferometers.\emph{\ Lect. Notes Phys. }\textbf{2001}, \emph{%
562, }403-438.

\bibitem{a19} Dimopoulos, S.; Graham, P. W.; Hogan, J. M.; Kasevich, M. A.
General relativistic effects in atom interferometry.\emph{\ Phys. Rev. D }%
\textbf{2008}, \emph{78}, 042003.

\bibitem{a20} Wicht, A.; L\"{a}mmerzahl, C.; Lorek, D.; Dittus, H.
Rovibrational quantum interferometers and gravitational waves. \emph{Phys.
Rev. A} \textbf{2008}, \emph{78}, 013610.

\bibitem{a21} Dubetsky, B. Optimization and error model for atom
interferometry technique to measure Newtonian gravitational constant.
http://arxiv.org/abs/1407.7287.

\bibitem{a21.1} Zorn, A.; Sonnad, V.; Libby, S. B.; Dubetsky, B.; and
Shverdin, M, A Semi-Classical, Closed Form Expression for the Phase Response
of a Vertical Symmetric Atomic Fountain, to be submitted to Phys. Rev. A,
2016.

\bibitem{a27.1} Higher order terms in the expansion of the potential have
also been considered in Ref. \cite{a27}.

\bibitem{c2.1} Wolf, P; Tourrenc, P. Gravimetry using atom interferometers:
Some systematic effects. \emph{Physics Letters A} \textbf{1999, }\emph{251,}
241-246.

\bibitem{c2} Bongs, K.; Launay, R.; Kasevich, M. A. High-order inertial
phase shifts for time-domain atom interferometers. \emph{Appl. Phys. B} 
\textbf{2006}, \emph{84}, 599-602.

\bibitem{c2.2} Peters, A.; Chung, K. Y.; Chu, S. High-precision gravity
measurements using atom interferometry. \emph{Metrologia} \textbf{2001}, 
\emph{38}, 25-61.

\bibitem{a36} Kasevich, M. A.; Dubetsky, B. Kinematic Sensors Employing Atom
Interferometer Phases. US Patent 2006/0249666 a1.

\bibitem{a22} Dickerson, S. M.; Hogan, J. M.; Sugarbaker, A.; Johnson, D. M.
S.; Kasevich, M. A. Multiaxis Inertial Sensing with Long-Time Point Source
Atom Interferometry. \emph{Phys. Rev. Lett.} \textbf{2013}, \emph{111},
083001.

\bibitem{a23} Beausoleil, R. G.; Hansch, T. W. Ultrahigh-resolution
two-photon optical Ramsey spectroscopy of an atomic fountain. \emph{Phys.
Rev. A} \textbf{1986}, \emph{33}, 1661-1670.

\bibitem{a24} Dubetsky, B. Gradiometer response on parallelepiped test mass.
Unpublished. \textbf{2008}.

\bibitem{a24a} For a test mass density$\rho ,$ the field $\delta g\ $is of
order $G\rho y_{\min }$. For inequality (\ref{2b}) to be satisfied, we must
have 
\begin{equation*}
\rho \ll \frac{M_{E}}{y_{\min }R_{E}^{2}}\approx 1.5\times 10^{12}\text{ kg/m%
}^{3},
\end{equation*}%
where $M_{E}$ and $R_{E}$ are Earth's mass and radius,respectively, and we
put $y_{\min }\sim 0.1$m. Since the right-hand-side (rhs) of this inequality
is 8 orders of magnitude larger than maximum atomic density existing on
Earth, assumption (\ref{2b}) is reasonable.

\bibitem{c1} Dubetsky, B.; Berman, P. R. Ground-state Ramsey fringes. \emph{%
Phys. Rev. A} \textbf{1997}, \emph{56}, R1091-R1094.

\bibitem{a25.1} See, for example, Landau L. D.; Lifshitz E. M. Statistical
Physics, Part 1. \emph{Pergamon Press, New York }\textbf{1980, }p. 20;
Klimontovich, Yu. L. Kinetic Theory of Nonideal Gases amd Nonideal Plasmas. 
\emph{Pergamon Press, New York }\textbf{1982, }Chapter 12.\emph{\ }

\bibitem{a25} Kol'chenko, A. P.; Rautian, S. G.; Sokolovskii, R. I.
Interaction of an atom with a strong electromagnetic field with the recoil
effect taken into consideration. \emph{Zh. Eksp. Teor. Fiz.} \textbf{1968}, 
\emph{55}, 1864-1873 [\emph{JETP} \textbf{1969}, \emph{28}, 986-990].

\bibitem{a25.2} Berman, P. R.; Malinovsky, V. S. Prinsiples of laser
spectroscopy and quantum optics. \emph{Prinston University Press, Prinston }%
\textbf{2011, Sec. 18.5}

\bibitem{a26} Giese, E.; Zeller, W.; Kleinert, S.; Meister, M.; Tamma, V.;
Roura, A.; Schleich, W. P. The interface of gravity and quantum mechanics
illuminated by Wigner phase space Atom Interferometry. In \emph{Proceedings
of the International School of Physics "Enrico Fermi";} Tino, G. N.;
Kasevich, M. Eds.; IOS Press; \textbf{2014}; \emph{188, }pp. 171 - 236.

\bibitem{a27} Hogan, J. M. Testing Gravity with Atom Interferometry.
www-conf.slac.stanford.edu/ssi/2011/Hogan-080311.pdf\emph{,} Talk in \emph{%
39th SLAC Summer Institute; }2011.

\bibitem{a28} Giltner, D. M.; McGowan, R. W.; Lee, S. A. Theoretical and
experimental study of the Bragg scattering of atoms from a standing light
wave. \emph{Phys. Rev. A }\textbf{1995}, \emph{52}, 3966-3972.

\bibitem{a29} Berman, P. R.; Dubetsky, B.; Cohen, J. L. High-resolution
amplitude and phase gratings in atom optics. \emph{Phys. Rev. A }\textbf{1998%
}, \emph{58}, 4801-4810.

\bibitem{a30} Dubetsky, B.; Berman, P. R. $\lambda $/4, $\lambda $/8, and
Higher Order Atom Gratings via Raman Transitions. \emph{Laser Physics} 
\textbf{200}2, \emph{12}, 1161-1170.

\bibitem{a31} Turlapov, A.; Tonyushkin, A.; Sleator, T. Talbot-Lau effect
for atomic de Broglie waves manipulated with light. \emph{Phys. Rev. A} 
\textbf{2005}, \emph{71}, 043612.

\bibitem{a32} L\'{e}v\`{e}que, T.; Gauguet, A.; Michaud, F.; Pereira Dos
Santos, F.; Landragin, A. Enhancing the Area of a Raman Atom Interferometer
Using a Versatile Double-Diffraction Technique. \emph{Phys. Rev. Lett.} 
\textbf{2009}, \emph{103}, 080405.

\bibitem{a33} Chiow, S.; Kovachy, T.; Chien, H.; Kasevich, M. A. 102$\hbar $%
k Large Area Atom Interferometers. \emph{Phys. Rev. Lett. }\textbf{2011}, 
\emph{107}, 130403.

\bibitem{a33.1} Kovachy, T.; Asenbaum P.; Overstreet1, C.; Donnelly, C. A.;
Dickerson, S. M.; Sugarbaker, A.; Hogan, J. M.; Kasevich M. A. Quantum
superposition at the half-metre scale. \emph{Nature} \textbf{2015, }\emph{528%
}\textbf{, }530-533\textbf{.}

\bibitem{a32.1} Dubetsky, B.; Berman, P. R. $\lambda /$8-period optical
potentials. \emph{Phys. Rev. A} \textbf{2002}, \emph{66}, 045402.

\bibitem{a34} Zorn, A. Unpublished. \textbf{2011}.

\bibitem{a35} Chemical elements listed by density.
http://www.lenntech.com/periodic-chart-elements/density.htm
\end{thebibliography}
\end{document}